\documentclass[fleqn,usenatbib]{mnras}

\usepackage[T1]{fontenc}

\DeclareRobustCommand{\VAN}[3]{#2}
\let\VANthebibliography\thebibliography
\def\thebibliography{\DeclareRobustCommand{\VAN}[3]{##3}\VANthebibliography}

\usepackage{graphicx}	
\usepackage{amsmath}	
\usepackage{amssymb}	
\usepackage{soul,xcolor}
\usepackage[normalem]{ulem}
\graphicspath{{./}{figures/}}
\newcommand\rst{\bgroup\markoverwith{\textcolor{red}{\rule[0.5ex]{2pt}{0.4pt}}}\ULon}

\DeclareMathOperator{\sech}{sech}

\title[The vertical structure of galactic discs]{The vertical structure of galactic discs: nonlocal gravity versus dark matter}

\author[T. Kashfi et al.]
{Tahere Kashfi,$^{1}$
Mahmood Roshan,$^{1,2}$\thanks{E-mail: mroshan@um.ac.ir}
\\
$^{1}$Department of Physics, Faculty of Science, Ferdowsi University of Mashhad, P.O. Box 1436, Mashhad, Iran\\
$^{2}$School of Astronomy, Institute for Research in Fundamental Sciences (IPM), 19395-5531, Tehran, Iran\\
}

\date{Accepted XXX. Received YYY; in original form ZZZ}

\pubyear{2023}

\begin{document}
\label{firstpage}
\pagerange{\pageref{firstpage}--\pageref{lastpage}}
\maketitle

\begin{abstract}
Recent isolated galactic simulations show that the morphology of galactic discs in modified gravity differs from that of the standard dark matter model. In this study, we focused on the vertical structure of galactic discs and compared the bending instability in the vertical direction for both paradigms. To achieve this, we utilized high-resolution N-body simulations to construct two models in a specific nonlocal gravity theory (NLG) and the standard dark matter model and compared their stability against the bending perturbations. Our numerical results demonstrate that the outer regions of the disc are more susceptible to the instability in NLG, whereas the disc embedded in the dark matter halo is more unstable in the central regions. We then interpret these results based on the dispersion relation of the bending waves. To do so, we presented an analytical study to derive the dispersion relation in NLG. Our numerical results align with the predictions of our analytical models. Consequently, we conclude that the analysis of bending instability in galactic discs offers an explanation for the distinct vertical structures observed in simulated galactic discs under these two theories. These findings represent a significant step towards distinguishing between the modified gravity and dark matter models.
\end{abstract}



\section{Introduction}\label{sec:intro}
Wide observational evidence indicates that many disc galaxies, including our host galaxy, show boxy or peanut-shaped structures in their central regions when they are seen edge-on \citep{Shaw1987, Weiland1994, Lutticke2000, Ciambur2017, Erwin2017, Kruk2019}. The presence of boxy/peanut bulges significantly affects the estimates of the stellar gravitational potential, forces, and orbital structures of their host galaxies \citep{Fragkoudi2015}. The study of the vertical structure of disc galaxies is thus an essential part of understanding and constraining models of galaxy formation and evolution.

The fastest secular evolution in the vertical direction is driven by buckling instability, leading to changes in bulge-to-disc ratio and other disc structural parameters \citep{Debattista2006}. This instability is the most violent thickening mechanism during which the symmetry about the mid-plane is broken and a distinct boxy/peanut-shaped structure forms in the inner part of the bar \citep{Martinez2004, Martinez2011, Erwin2016, Collier2020, Sellwood2020, Ciambur2021, Cuomo2023}

Buckling instability was detected in the numerical simulations of \citet{Combes1981} for the first time. However, after 20 years of numerical studies showing buckling instability, the nature of this phenomenon is still unclear. Some authors claim that buckling is due to the bending (fire-hose) instability \citep{Toomre1966, Araki1985, Raha1991, Merritt1991, Merritt1994, Debattista2006, Zana2019, Lokas2020, Collier2020} which leads to the breakup of the vertical symmetry and leaves a thickened bar. At the same time, the others relate it to the orbital resonance \citep{Combes1990, Pfenniger1991, Quillen2002, Valpuesta2006, Quillen2014, Saha2018, Li2023}. However, some studies argued that the latter mechanism could not fully explain buckling \citep{Merritt1994, Lokas2020}.

By far, numerical simulations are the main source of our information about the vertical evolution of galactic discs. N-body simulations conducted within the context of modified gravity theories imply that the vertical structure of disc galaxies has some obvious differences compared with the standard dark matter picture \citep{Tiret2007, Roshan2018, Roshan2019, Ghafourian2020, Roshan2021}. For example, discs are more warped and flared in modified gravity, and the peanut configurations are weaker in comparison with the standard dark matter model. It is necessary to mention that there are some modified gravity models that postulate the existence of dark matter particles and focus on dark energy. In this study, by modified gravity we mean a theory that denies the existence of dark matter particles. Specifically, examples include MOND \citep{Milgrom1983}, MOG \citep{Moffat2006}, and NLG \citep{Hehl2009a, Hehl2009b}.

One of the primary objective of this work is to differentiate between the dark matter model and the modified gravity theory under study. To be specific, we investigate the reason for the different vertical structures of galactic discs in the context of modified gravity and the standard dark matter model. The investigation of the underlying factors contributing to the disparate behaviours observed in these two viewpoints allows for a deeper understanding of how the specific properties of each model influence the evolution of galactic discs. Specifically, we study the unstable bending waves in the vertical direction through N-body simulations of isolated discs under the effect of a specific Non Local Gravity theory (NLG) and embedded in a dark matter halo. To do so, we first follow the analytical studies of bending instability in the stellar system and obtain a dispersion relation for bending waves in NLG. Then, we analyse the results of the numerical simulations based on the predictions of our analytical approach. We conclude that the unstable bending modes are different in these two paradigms, and the numerical results can be interpreted based on the dispersion relation of bending waves. 

The structure of this paper is as follows: Section \ref{NLG} provides a brief introduction of NLG. In Section \ref{sec:analys}, we present the analytical approach for the propagating of bending waves in NLG. In Section \ref{sec:simu}, we introduce isolated simulations of galactic discs in two gravitational contexts, purring emphasis on bending instability. Numerical results and their interpretations are performed in Section \ref{sec:result}. The final discussion and conclusions can be found in Section \ref{sec:con}.

\section{Non local gravity}\label{NLG}
Despite numerous experiments searching for dark matter particles, no direct evidence has been found so far, leading to ongoing uncertainty regarding the nature of dark matter. This has naturally led to the possibility that what is currently attributed to dark matter in astrophysics and cosmology may, in fact, be an aspect of the gravitational interaction that is not yet fully understood or accounted for within our current theoretical framework.

NLG is a modified gravity which hypothesize that gravitation possesses a history-dependent nature, much like nonlocal electrodynamics in media. This suggests that the gravitational interaction exhibits a form of nonlocality, wherein there is an influence or ``memory'' from past events that persists over time. In NLG, there is a perspective that the phenomena traditionally attributed to dark matter in astrophysics and cosmology could potentially be explained by the nonlocal aspect of gravitational interactions. This viewpoint offers an alternative approach to understanding the observed effects without invoking the existence of dark matter. However, it is essential to note that at present, the nonlocal character of gravity in NLG has not been claimed to completely replace the need for dark matter on all physical scales. For instance, NLG has not been applied to well-known systems such as the Bullet cluster, where the observations suggest the presence of cold dark matter particles. Similarly, the cosmic structure formation where the cold dark matter is the main player has not been thoroughly investigated in the context of NLG. Therefore, it remains uncertain whether NLG can accurately reproduce the observed cosmic microwave background and matter power spectra on linear cosmological scales. It is important to emphasize that NLG is still in its early stages of development, and only some of its implications have been tested against observational data \citep{Mashhoon2017}.

This tetrad theory established upon the frame field of a fundamental family of observers in space-time. The nonlocal generalization of Einstein’s field equation is given by

\begin{equation}\label{eq4}
	G_{\mu\nu}+\mathcal{N}_{\mu\nu}=\kappa T_{\mu\nu}-\Lambda \ g_{\mu\nu}+\ Q_{\mu\nu},
\end{equation}
where $\Lambda$ is the cosmological constant and $\mathcal{N}_{\mu\nu}$ and $Q_{\mu\nu}$ are two tensors which are given by 
\begin{equation}\label{eq5}
	\mathcal{N}_{\mu\nu}=g_{\nu\alpha}\ e_{\mu}^{\ \ \hat{\gamma}}\frac{1}{\sqrt{-g}}\frac{\partial}{\partial x^{\beta}}\Big(\sqrt{-g}\ N^{\alpha\beta}_{\ \ \ \hat{\gamma}}\Big),
\end{equation}

and
\begin{equation}\label{eq6}
	Q_{\mu\nu}=C_{\mu\rho\sigma}N_{\nu}^{\ \ \rho\sigma}-\frac{1}{4} g_{\mu\nu}C_{\delta\rho\sigma}N^{\delta\rho\sigma}.
\end{equation}

Here, $C_{\mu\rho\sigma}$ is the torsion tensor and $N_{\nu\rho\sigma}$ is the nonlocality tensor  which is an average over the past history of the gravitational field. This tensor includes a causal scalar kernel that determines the nonlocal deviation from the General Relativity in this theory. This kernel is in general a function of the coordinate invariants, but the exact form of it is still unknown. For the exact form and derivation of these tensors, we refer the reader to \citet{Mashhoon2017}. It worth mentioning that NLG is still in its early stages of development, and apart from the Minkowski space-time, no exact solution to the field equation of nonlocal gravity has been discovered thus far. However, some cosmological implications of this theory have been recently investigated \citep{Tabatabaei2022, Tabatabaei2023c, Tabatabaei2023b, Tabatabaei2023a}. 

In order to explore the consequences of NLG on galactic scales, we need the weak-field limit of the field equations. At this limit, the field equations of NLG are reduced to \citep{Mashhoon2017}
\begin{equation}\label{eq7}
	\nabla^2\Phi(\mathbf{x})=4\pi G\rho(\mathbf{x})+4\pi G\int q(\mathbf{x}-\mathbf{x}')\rho(\mathbf{x'})d^3x', 
\end{equation}

The nonlocal aspect of gravity appears as an extra source of matter in equation \eqref{eq7}. The density of this extra source is the convolution of the reciprocal kernel $q$ with the density of matter $\rho(\mathbf{x})$,   
\begin{equation}\label{eq8}
	\rho_D(\mathbf{x})=\int q(\mathbf{x}-\mathbf{x}')\rho(\mathbf{x'})d^3x'.
\end{equation}

In this sense, the nonlocal aspect of gravity has the interpretation of the density of effective dark matter, $\rho_D$, which the existence and distribution of this density, directly depends on the distribution of the baryonic matter density. It means that there is no effective dark matter in the complete absence of baryonic matter. Using the modified virial theorem it has been shown that $\rho+\rho_D$ is the density of the effective dynamical mass and the nonlocal aspect of gravity can simulate the role of dark matter halos in galactic scales \citep{Mashhoon2017}. So, galaxies in  NLG are supposed to be surrounded by effective dark matter halos which keep the rotation curves flat and control the dynamics of their periphery.

Certain characteristics of the effective dark matter density, $\rho_D$, exhibit distinguishing features when compared to the density profile of the standard dark matter model. In the NLG model, the effective dark matter component aligns with the distribution of baryonic matter. Notably, in spiral galaxies, the spiral patterns appear in the the corresponding effective dark matter as well. While this association is absent in the conventional dark matter framework. This disparity in behaviour has direct implications for the secular evolution of the galactic disc, influencing both its vertical and radial dynamics. Moreover, in contrast to the standard dark matter model, the density function $\rho_D(\mathbf{x})$ exhibits a characteristic smoothness and lacks cusps. Additionally, the overall abundance of effective dark matter within dwarf galaxies is typically lower than what is anticipated within the dark matter paradigm \citep{Roshan2022ApJ}.

In the Newtonian limit, two possible forms of the reciprocal kernel $q(\mathbf{x}-\mathbf{x'})$ are given by \citep{Mashhoon2017},
\begin{equation}\label{eq9}
	q_1=\frac{1}{4\pi \lambda_0}\frac{1+\mu_0(a_0+r)}{r(a_0+r)}e^{-\mu_0 r},
\end{equation}

and
\begin{equation}\label{eq10}
	q_2=\frac{1}{4\pi \lambda_0}\frac{1+\mu_0(a_0+r)}{(a_0+r)^2}e^{-\mu_0 r},
\end{equation}

where $r=|\mathbf{x}-\mathbf{x}'|$. Here $a_0$, $\lambda_0$ and $\mu_0$ are three positive free parameters which control the behaviour of the kernel in different scales. $\lambda_0$ is the nonlocal parameter and expected to be smaller than the large-distance parameter ($\mu^{-1}_0$) and larger than the short-distance parameter ($a_0$). 

On galactic scales, it is anticipated that the reciprocal kernel $q$ can be determined by analyzing the rotation curves of spiral galaxies. By comparing the predictions of NLG with the observed rotation curves, the aim is to determine the values of $a_0$ , $\lambda_0$, and $\mu_0$ that yield the best agreement with the data. This iterative fitting procedure involves adjusting these parameters to minimize any discrepancies between the observed rotation curves and the predictions of NLG.

In the galactic scales, the short-distance behaviour is unimportant, so we set $a_0=0$. In this way, $q_1$ and $q_2$ both reduce to the following relation
\begin{equation}\label{eq11}
	q=\frac{\alpha_0\mu_0}{8\pi}\frac{(1+\mu_0 r) e^{-\mu_0 r}}{r^2},
\end{equation}
where $\alpha_0$ is defined as $\alpha_0=2/\lambda_0\mu_0$. The optimal values for the parameters $\alpha_0$ and $\mu_0$ of the reciprocal kernel $q$ are derived by analyzing a subsample of galaxies from THINGS catalog. The best fit values of these parameters are $\alpha_0=10.94\pm 2.56$ and $\mu_0=0.059\pm 0.028\ {\rm kpc^{-1}}$. Remarkably, when these parameters of the NLG model are utilized to determine the rotation velocity, the calculated values of $V_{\text{flat}}$ align with the Tully-Fisher relation. Moreover, by extending the application of NLG to clusters of galaxies, it has been demonstrated that the dynamics of clusters are in agreement with the observed baryonic content \citep{Rahvar2014}.

It is important to note that the same range of free parameters works properly for some ultra-diffuse galaxies \citep{Roshan2022ApJ}. Furthermore, \citet{Zhoolideh2017} demonstrated that when employing these specific values for the free parameters, NLG can successfully match the rotation curves of dwarf galaxies. However, this fitting comes at the cost of having consistently high stellar mass-to-light ratios at a wavelength of $3.6\ \mu m$.

Using the modified Poisson's equation \eqref{eq7}, it is straightforward to find the gravitational field due to a point mass. Therefore, the gravitational force on a point mass $m$ due to point mass $m'$ is given by
\begin{equation}\label{eq12}
	\mathbf{F}_{\text{NLG}}(\mathbf{x})=-Gmm'\frac{\mathbf{x}-\mathbf{x'}}{|\mathbf{x}-\mathbf{x'}|^3}\Bigg[1+\int_0^{|\mathbf{x}-\mathbf{x'}|} 4\pi s^2 q(s) ds\Bigg],
\end{equation}
where the deviation from Newton's law is due to the spherical distribution of the effective dark matter density surrounding $m'$. It should be noted that based on Newton's shell theorem, $m$ is only affected by the portion of effective dark matter which is distributed within a sphere of radius $r=|\mathbf{x}-\mathbf{x'}|$ \citep{Roshan2022}.

The kernel of the theory in galactic scales is given by equation \eqref{eq11}. So equation \eqref{eq12} reduces to
\begin{equation}\label{eq13}
	\mathbf{F}_{\text{NLG}}(\mathbf{r})=-Gmm'\frac{\mathbf{r}}{r^3}\Bigg[1+\frac{\alpha_0 \mu_0 r}{2}\bar{U}(\mu_0 r)\Bigg],
\end{equation}
where
\begin{equation}\label{eq14}
	\bar{U}(x)=\frac{2e^{-x}}{x}\big(e^x-1-\frac{x}{2}\big).
\end{equation}
This function starts from $\bar{U}=1$ at $x=0$ and then decreases monotonically as $x$ increases. As $x\rightarrow \infty$, $\bar{U}(x)$ tends to zero.

It should be mentioned that, this force is conservative, adheres to Newton's third law of motion, and always possesses an attractive nature at all. It can be decomposed into two distinct components: an enhanced ``Newtonian'' component and a repulsive ``Yukawa'' component. The exponential decay in the Yukawa term stems from the phenomenon of spatial gravitational memory gradually fading away in an exponential manner as the distance between objects increases \citep{Mashhoon2017}.

\section{Bending instability in NLG}\label{sec:analys}
As mentioned in the introduction section, the vertical structure of simulated discs appears to be different in the modified gravity and dark matter models. The peanuts are weaker in modified gravity, while the outer parts of the discs are thicker. In this section, we present an analytic description for the propagation of bending waves and derive the dispersion relation in the NLG framework. Our aim is to explain the different vertical structures observed in simulated galactic discs based on the stability of the bending waves in two theories. Examining the stability of the dispersion relation and analysing its constituent terms will provide valuable insights for the interpretation of the subsequent numerical results.

The stability of bending waves in a stellar system has been investigated by several authors \citep{Toomre1966, Fridman1984, Araki1985, Merritt1991, Merritt1994, Sellwood1996}. However, one of the seminal analytical approaches was obtained by \citet{Toomre1966}. 

Toomre considered an infinitely thin slab of stars with a uniform surface density $\Sigma$ and Gaussian distributions of velocity both parallel to the plane and perpendicular to it. He showed that the frequency, $\omega$ is related to the wave number $k$ of the bending waves through the following dispersion relation;
\begin{equation}\label{eq1}
\omega^2=2\pi G\Sigma|k|-\sigma^2 k^2,
\end{equation}

where $\sigma$ is the velocity dispersion in a given horizontal direction.

Considering that the slab's thickness is of order $\sigma_z^2/G\Sigma$, such a system is stable against the unstable bending modes if the ratio of the vertical dispersion velocity $\sigma_z$ to the horizontal velocity dispersion $\sigma$  exceeds some critical value which \citet{Toomre1966} estimated to be $0.30$. 

Toomre derived the dispersion relation \eqref{eq1} for an isolated disc. Therefore, in the presence of some external potentials like a dark matter halo, the extra terms corresponding to the additional restoring forces should be included as follows \citep{Sellwood1998, Binney2008},
\begin{equation}\label{eq2}
\omega^2=\nu^2_{\text{ext}}+2\pi G\Sigma|k|-\sigma^2 k^2,
\end{equation}

where
\begin{equation}\label{eq3}
\nu^2_{\text{ext}}=\Big|\frac{\partial^2\Phi_{\text{ext}}}{\partial z^2}\Big|_{z=0}.
\end{equation}

Here $\Phi_{\text{ext}}$ is the potential due to all galaxy components other than the disc. In this relation, the vertical frequency due to the external potentials and the self-gravity of the disc tend to stabilize the bending waves, while the horizontal dispersion velocity has a destabilising effect.

Although galaxies are not razor-thin discs, this approximation gives an insight into the stability of bending waves in real discs. In the following, we derive the dispersion relation for the bending perturbations in the context of NLG. 

To do so, we follow Toomre's approximation and examine the behaviour of a razor-thin sheet with a surface density $\Sigma$, under the influence of bending perturbations. We assume that the sheet is initially located in the plane $z=0$, and the stars have Maxwellian velocity distribution
\begin{equation}\label{eq15}
f(v_x,v_y)\propto e^{\frac{1}{2}\frac{(v^2_x+v^2_y)}{\sigma^2}}.
\end{equation}

Now we consider a small perturbation in the form of a planar wave traveling in the $x$-direction
\begin{equation}\label{eq16}
z(x,t)=z_0 e^{i(kx-\omega t)}.
\end{equation}

It should be noted that in a razor-thin sheet, the self-gravity of the sheet is strong enough to locally constrains stars to move together in their vertical displacements. So the sheet bends coherently, and the bending can be completely described in terms of a single function $z(x,t)$ \citep{Merritt1994, Binney2008, Sellwood2013}.

Since the bending wave \eqref{eq16} travels in the $x$-direction, we can neglect the motion of stars in the $y$-direction because this motion has no effect on the interaction of the stars with the wave. So the vertical acceleration of a star, as it follows the bend, is given by
\begin{equation}\label{eq17}
\frac{d^2z(x,t)}{dt^2}=\Big(\frac{\partial^2}{\partial t^2}+2v_x\frac{\partial^2}{\partial x\partial t}+v^2_x\frac{\partial^2}{\partial x^2}\Big)z(x,t).
\end{equation}

Here we replaced the Lagrangian derivative with the relation $d/dt=\partial/\partial t+v_x\partial/\partial x$. By averaging over all of the stars at a given location, we get
\begin{equation}\label{eq18}
\frac{d^2z(x,t)}{dt^2}=\Big(\frac{\partial^2}{\partial t^2}+\sigma^2\frac{\partial^2}{\partial x^2}\Big)z(x,t),
\end{equation}

where we used $\langle v_x\rangle=0$ and $\langle v^2_x\rangle=\sigma^2$.

Equation \eqref{eq18} must be equal to the vertical acceleration due to the self-gravity of the sheet $g_z(x,t)$ and the gravitational restoring force per unit mass caused by the external potential. Thus, the equation of motion for a star as it moves around the bend can be expressed as
\begin{equation}\label{eq19}
\Big(\frac{\partial^2}{\partial t^2}+\sigma^2\frac{\partial^2}{\partial x^2}\Big)z(x,t)=g_z(x,t)-\nu^2_{\text{NLG}}z(x,t).
\end{equation}

Here $g_z(x,t)$ is determined only by the Newtonian potential of the sheet, which means that the nonlocal aspect of gravity does not contribute in this term. So we can use the same equation for a Newtonian sheet derived in \citet{Binney2008} (see also \citet{Merritt1994}), namely
\begin{equation}\label{eq20}
g_z(x,t)=-2\pi G\Sigma |k|z(x,t).
\end{equation}

The second term in the right-hand side of equation \eqref{eq19} denotes the acceleration due to the resorting gravitational force. Since our sheet is under the effect of external potential caused by the effective dark matter density, only the nonlocal aspect of gravity should be included in this term. So, the vertical frequency $\nu^2_{\text{NLG}}$, is determined by the nonlocal terms in equation \eqref{eq13}.

By substituting equations \eqref{eq16} and \eqref{eq20} in the equation of motion \eqref{eq19}, we obtain the dispersion relation for the bending wave in NLG, 
\begin{equation}\label{eq22}
\omega^2=\nu^2_{\text{NLG}}+2\pi G\Sigma |k|-\sigma^2 k^2.
\end{equation}

In this relation, the vertical frequency $\nu^2_{\text{NLG}}$ and the self-gravity of the sheet have a stabilizing effect, while the horizontal dispersion destabilises the sheet against bending perturbations.

In the following sections, we present our analysis of bending instability in the N-body simulations of galactic discs in NLG and a dark matter model. We interpret the results of our simulations using the analytic dispersion relations \eqref{eq2} and \eqref{eq22} and show that this instability can be responsible for the different morphology of the discs in the vertical direction of our models.   
\section{Numerical Simulation}\label{sec:simu}
To compare the bending instability in a disc embedded in a dark matter halo or under the effect of nonlocal gravity, we use high-resolution N-body simulations. We construct two models and evolve them using the \textit{GALAXY} code \citep{Sellwood2014}\footnote{http://www.physics.rutgers.edu/galaxy}. \textit{GALAXY} is a widely-used N-body code and robust tool for simulating isolated galaxies. It offers the capability to generate equilibrium initial conditions for various combinations of halos, bulges, and discs, encompassing a wide range of density profiles commonly used in galaxy simulations. With eleven different methods available, users can select the most suitable approach to compute the gravitational field of the particles. It is also provides different 3D and 2D grids for many applications of interest. Once the user sets the numerical parameters, the code generates a file containing the initial positions, velocities, and, if desired, masses of the particles. By reading this file, the code calculates the gravitational accelerations acting on each particle and performs numerical integration to update their velocities and positions over short time steps. These steps can be repeated as needed, and the code allows for saving the final state, enabling easy resumption of the integration.
	
In the case of the dark matter model, our simulation implements two components; a spherical dark matter halo and a stellar disc. Consequently, we used a hybrid grid constructed from a three-dimensional cylindrical grid for the disc particles and a spherical grid for the halo. 

On the other hand, to conduct galactic simulations in NLG, we have modified the \textit{GALAXY} code to include nonlocal corrections to the gravitational force; see \citet{Roshan2021} for more detail. It is necessary to mention that in this case, there is only one mass component, e.g., the baryonic stellar disc. Therefore, we use a single three-dimensional cylindrical grid.
\subsection{Initial conditions}
In the following section, we provide a detailed description of how the initial conditions for these simulations are established.

The models under study are flat discs with an exponential radial density profile given by
\begin{equation}\label{eq23}
\rho(R,z)=\frac{M_d}{4\pi z_d R_d^2} e^{-\frac{R}{R_d}} \text{sech}^2\left(\frac{z}{2z_d}\right)
\end{equation}

where $M_d$ is the disc mass and $R_d$ and $z_d$ are the length scale and height scale of the disc, respectively. 

In Newtonian discs, the $\sech^2$ vertical density profile is a well-known solution of the vertical Jeans equation combined with Poisson's equation \citep{Spitzer1942}. In our study, we used this profile as an approximate initial density for the \textit{GALAXY} code to create a reasonable equilibrium state for the disc. The code integrates the vertical Jeans equation for the slab by neglecting radial variation, which yields the following relation for the vertical dispersion velocity,
\begin{equation}\label{eq23-1}
	\sigma^2_z(R,z)=\frac{1}{\rho(R,z)}\int^{\infty}_z\rho(R,z')\frac{\partial\Phi}{\partial z'}dz'.
\end{equation}

This method generally yields a better equilibrium state than the isothermal $\sech^2$ profile. It is important to note that in NLG simulation, the total potential $\Phi$ in equation \eqref{eq23-1} is replaced with the solution of the modified Poisson's equation \eqref{eq7}.

For the dark matter simulation, we employ a live Plummer halo (LPH) with the following density profile;
\begin{equation}\label{eq24}
	\rho_{\text{LPH}}(r)=\frac{3 M_h}{4\pi r_h^3}\Bigg[1+\Big(\frac{r}{r_h}\Big)^2\Bigg]^{-5/2},
\end{equation}
where $M_h$ is the halo mass and $r_h$ is the halo length scale.

As mentioned before, we use a hybrid grid for dark matter simulation. The grid size selected for our spherical mesh is $1001\times 2501$, where the first number indicates the number of radial grid shells and the second one is the radius of the grid's outer boundary. For the three-dimensional cylindrical grid employed for the disc, the mesh size is $193\times 224\times 81$. These numbers indicate the number of mesh points in the radial, azimuthal, and vertical directions. Table \ref{tab1} briefly highlights some key parameters of these simulations. 

\begin{table*}
	\centering
	\caption{Simulations parameters. The columns contain (1): Acronym used for simulation models (2): the initial disc mass in units of $10^{10}\ M_{\odot}$, (3): the radial length scale (kpc), (4): the vertical height scale (kpc), (5)-(6): the free parameters of the NLG model, (7): the halo mass scaled by $M_d$, (8): the halo length in terms of $R_d$, (9): the basic time step in units of $\tau_0=4.2$ Myr, (10): number of rings, spokes, and planes in the cylindrical polar grid, (11): the number of shells in the spherical grid.} \label{tab1}
	\begin{tabular}{ccccccccccc} 
		\hline
		\hline
		Simulation & $M_d$ & $R_d$ & $z_d$ & $\alpha_0$ & $\mu_0$ & $M_h/M_d$ & $r_h/R_d$ & $\delta t$ & Cylindrical polar grid & Spherical grid\\
		
		Model & ($10^{10}M_{\odot}$) & (kpc) & (kpc) &  & ($1/R_d$) &\\
		\hline
		NLG & 1.8 & 1.13 & 0.09 & 10.94 & 0.031 & -- & -- & 0.01 & $193\times 224\times 81$ & --\\
		LPH & 1.8 & 1.13 & 0.09 & -- & -- & 2.98 & 11.2 & 0.01 & $193\times 224\times 81$ & 1001\\
		\hline
	\end{tabular}
\end{table*}

In these simulations, our objective is to establish a system characterized by rotational dominance. As such, the rotational velocity is determined through a combination of the mass distribution and the gravitational law. It proves useful to implement the same initial conditions for the baryonic disc in all models. In this way, one can make a meaningful comparison between the different models. To do so, we changed the free parameters of the dark matter halo density and the NLG model to construct almost the same rotation curves. It is worth noting that the chosen values of the free parameters in the NLG model, are consistent with the observational constraints introduced in \citet{Rahvar2014}. 

\begin{figure}
	\centering
	\includegraphics[width=1.0\linewidth]{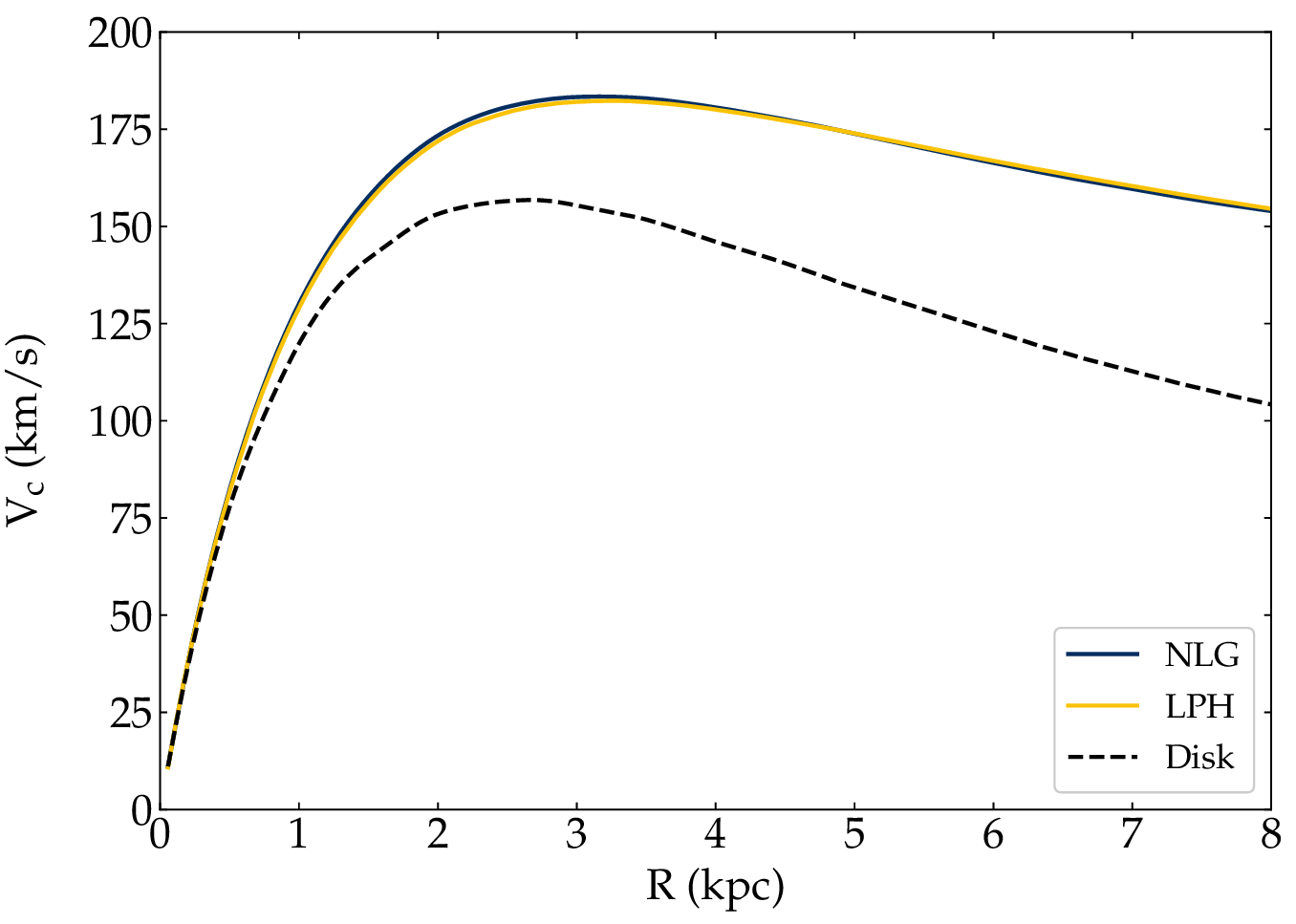}
	\caption{Initial rotational velocities for NLG (blue) and LPH (yellow) model. The black dashed lines indicate the Newtonian contribution of the disc to the rotational velocity. \label{fig:RC}}
\end{figure}

Figure.~\ref{fig:RC} illustrates the best fit of the rotation curves. It is important to note that our selection of disc parameters is specifically adopted to produce a consistent maximal disc configuration. This deliberate choice ensures the occurrence of bar formation and buckling instability within our simulations.

To avoid the local instabilities within the disc, the radial velocity dispersion is determined based on the Toomre criterion \citep{Toomre1964}. That is, the disc particles are assigned equilibrium orbital velocities, incorporating an initial radial velocity dispersion $\sigma_R$, such that $Q=1.5$ across all radii, where $Q$ is the Toomre parameter defined as
\begin{equation}\label{eq25}
	Q=\frac{\sigma_R \kappa}{3.36 G \Sigma},
\end{equation}
in which, $\kappa$ is the epicyclic frequency in the disc mid-plane, and $\Sigma$ is the surface density. Basically, in this way the radial velocity distribution is determined. The determination of the azimuthal velocity dispersion is achieved by solving the Jeans equations. The whole above mentioned process is primarily carried out by the program \texttt{smooth} in the \textit{GALAXY} code \citep{Sellwood2014}.

\begin{figure}
	\centering
	\includegraphics[width=1.0\linewidth]{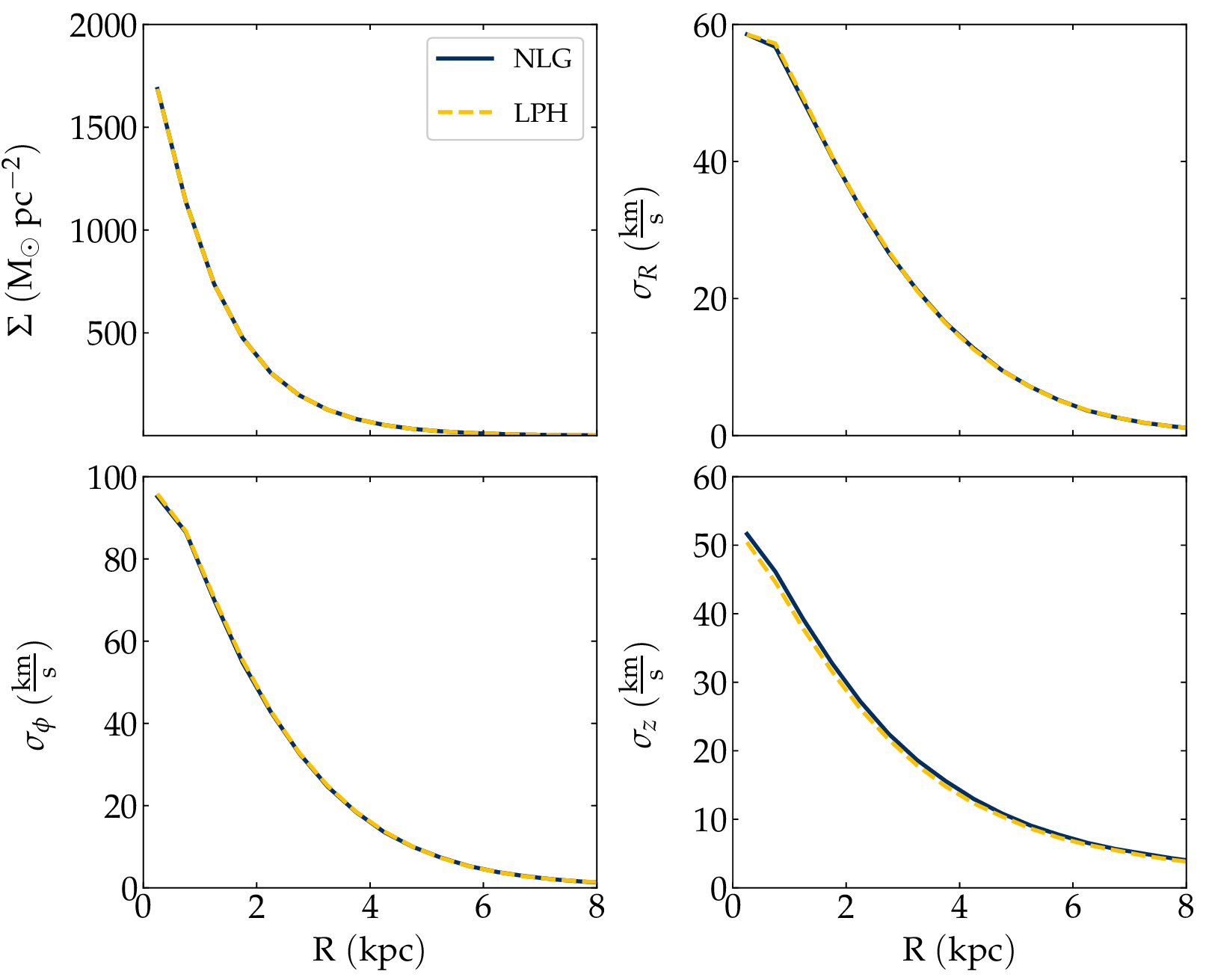}
	\caption{\textit{Top left}: the initial surface density, shown for NLG (blue) and LPH (yellow) models. \textit{Other panels}: the initial velocity dispersion profiles of our models. \label{fig:Den-dis}}
\end{figure}

To ensure that the internal properties, such as velocity dispersions, are consistent between different models, we use Figure~\ref{fig:Den-dis} to compare the initial surface density $\Sigma$ and the radial, azimuthal, and vertical velocity dispersions ($\sigma_R$, $\sigma_\phi$, and $\sigma_z$, respectively) of our models. It should be mentioned that these quantities are evaluated at the mid-plane of the disc. Specifically, we restricted our analysis to particles confined within $|z|\leq 1$ kpc, and projected all these particles onto the $z=0$ plane. The figure shows that the velocity dispersions of both models are appropriately consistent. Since, we attempted to create initial disc models with identical rotation curves, the distribution of the effective dark matter density in NLG and the dark matter halo are somewhat similar. Therefore, we can expect comparable internal characteristics for the discs. However, it is important to note that the effective dark matter in NLG is not spherical, which may account for the slight difference observed in the initial distribution of $\sigma_z$. Despite this difference, the overall consistency of the velocity dispersions demonstrates that our simulations start from similar initial conditions.

We have run our simulations in the units where $G=M_d=R_d=1$. So the time unit can be expressed as $\tau_0=(R_d^3/G M_d)^{1/2}$, which is equal to $\tau_0=4.2$ Myr for our models. We employed a total of $N=1\times 10^6$ equal mass particles to represent the stellar disc. Additionally, for the dark matter simulation, we utilized $10^6$ particles for the halo. The evolution of our models is observed and analyzed over a period of 5 Gyr.
\section{Results}\label{sec:result}

\subsection{The onset of bending instability}\label{buckling_time}

The initial indication of unstable bending modes is observed through the buckling of the bar structure. This buckling event subsequently initiates the propagation of bending waves throughout the disc \citep{Khoperskov2019}. Consequently, to ascertain the onset of the bending instability, it becomes imperative to investigate the time at which the buckling occurs.
  
The buckling instability results in weakening and thickening the bar. After the bar forms in the disc, the rapid streaming motions of the stars, which are trapped in the bar, increase the horizontal velocity dispersion more and more until the buckling instability occurs \citep{Collier2020}. This process manifests as a sudden decrease in the bar strength.  

In order to detect the buckling time, we compute the time evolution of the buckling amplitude $A_{\text{buck}}$. This quantity is defined as the following expression \citep{Debattista2006, Debattista2020, Anderson2022MNRAS}
\begin{equation}\label{eq26}
A_{\text{buck}}(R)=\frac{1}{M(R)}\Big|\sum^N_{k=1} z_k m_k e^{(2i\phi_k)}\Big|
\end{equation}	
where $z_k$ is the $z$ coordinate of $k$th particle, $\phi_k$ is the azimuthal angle, and $m_k$ is the particle mass. We divide the disc into annuli of equal width and consider only particles with $|z|\leq 10 z_d$ to avoid the discreteness noise caused by the particles which have higher vertical distances. The total mass of the particles in the annulus with radius $R$ is denoted by $M(R)$. Then we compute $A_{\text{buck}}$ in each radial bin and consider the maximum value of $A_{\text{buck}}(R)$ as the buckling amplitude for each snapshot.

\begin{figure}
	\centering
	\includegraphics[width=1.0\linewidth]{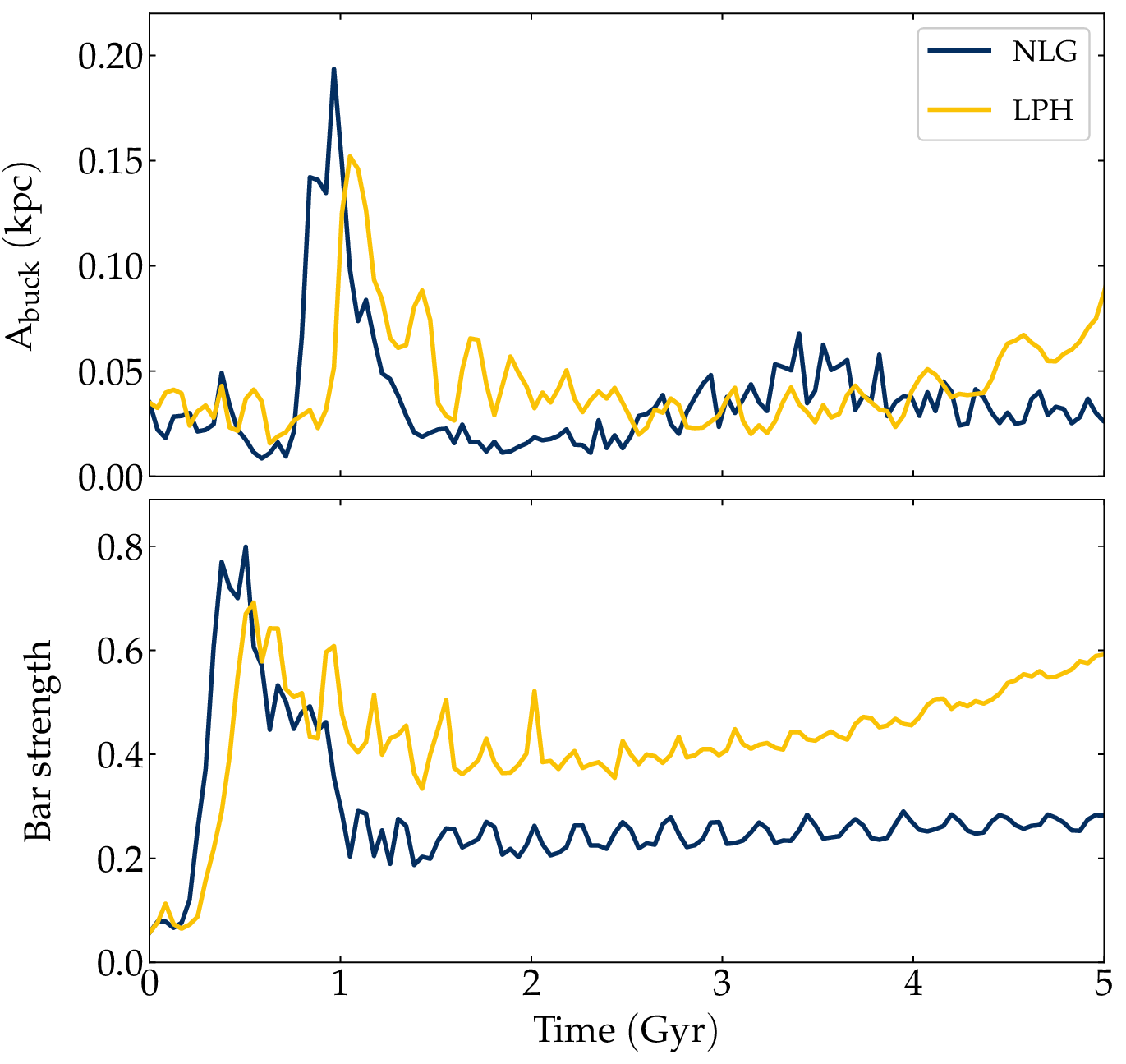}
	\caption{\textit{Top}: time evolution of the buckling amplitude, $A_{\text{buck}}$, for NLG (blue) and LPH (yellow) model. \textit{Bottom}: the evolution of bar strength in our models. A sudden decrease in the bar strength occurs almost simultaneously with an increase in $A_{\text{buck}}$.\label{fig:buckling}}
\end{figure}

The top panel in Figure~\ref{fig:buckling} shows the time evolution of buckling amplitude for the NLG and LPH models. We also plotted the time evolution of the bar strength in the bottom panel of this figure. This quantity is defined as
\begin{equation}\label{eq27}
	A_2^{\text{max}}=\max[A_2(R)],
\end{equation}
where $A_2(R)$ is the Fourier amplitude of mode $m=2$ at radius $R$. The amplitude of the $m$th Fourier mode is given by,
\begin{equation}\label{eq28}
	A_m(R) = \frac{1}{M(R)}\Big|\sum^N_{k=1} m_k e^{(i m\phi_k)}\Big|.
\end{equation}

From this figure, one can see that the buckling instability occurs at $t\simeq0.76$ Gyr in the NLG model, while in the LPH model, it happens later at about $t\simeq0.92$ Gyr. 

We see the common feature that buckling instability happens after the bar instability. As we expected, the bar instability develops more rapidly in NLG compared to the LPH model. It should be mentioned that this finding is independent of the adopted disc/halo profiles as demonstrated by \citet{Ghafourian2020}. They studied the evolution of a Kuzmin-Tommre disc with three different dark matter halos and found that the differences in bar evolution between modified gravity and dark matter models remained unchanged when they varied the initial mass densities. Interestingly, similar behaviour has been reported in MOND by different groups \citep{Tiret2007, Roshan2021}.

Also, for these maximal disc simulations, the final magnitude of the bar is weaker in NLG compared to the dark matter model LPH. Along the same line, \citet{Roshan2019}, \citet{Roshan2021} found that the bars are weaker not only in NLG but also in MOND.

\subsection{Effective dark matter density}
There are two quantities in the dispersion relation which control the stability of the disc at different radii. Both of these quantities, namely the vertical frequency $\nu$ and the velocity dispersion $\sigma$, are directly influenced by the distribution of the effective dark matter in the NLG model, and by the dark matter density in the LPH model. Therefore, the main differences that appear in the vertical structure of the discs, is related to the different distribution of the effective dark matter and the particle dark matter. So it is instructive to examine the evolution of these components in our models.

From equation \eqref{eq8} one can see that the distribution of effective dark matter in NLG, is closely connected with the density of matter. So, one expects that the distribution of dark matter halo in the LPH model to be different from the effective dark matter density in the NLG model. For an example, notice that there are empty space between dark matter particles in the LPH model; whereas the effective dark matter in the NLG model exists everywhere in the system. Of course, it appears with higher density where the baryonic matter density is higher. Furthermore, the particle dark matter have a spherical distribution while the effective dark matter follows the symmetry properties of the baryonic matter. In our system, we expect that the effective dark matter has azimuthal symmetry.

\begin{figure*}
	\centering
	\includegraphics[width=0.5\linewidth]{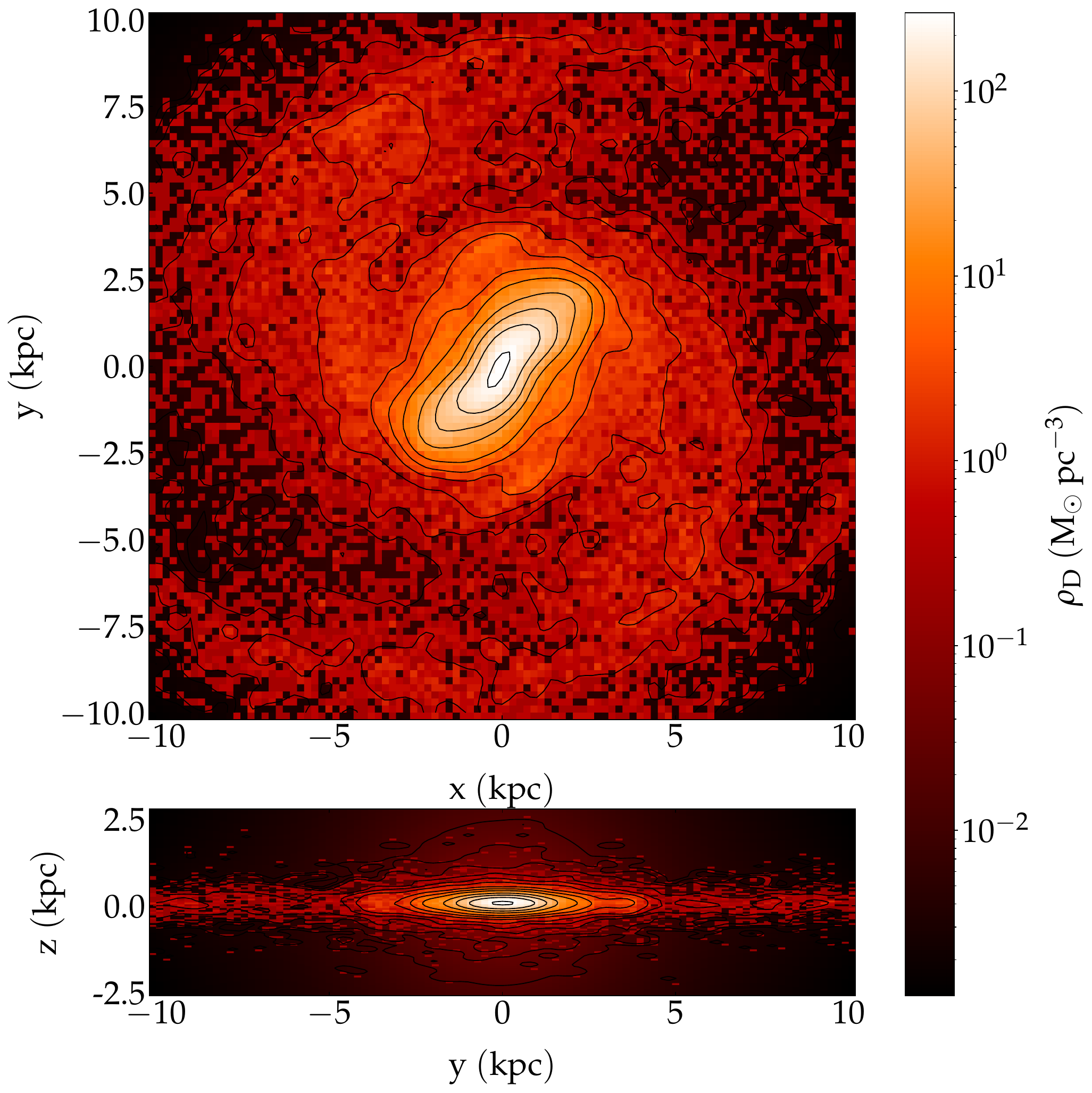}\hspace*{0.25cm}	
	\includegraphics[width=0.5\linewidth]{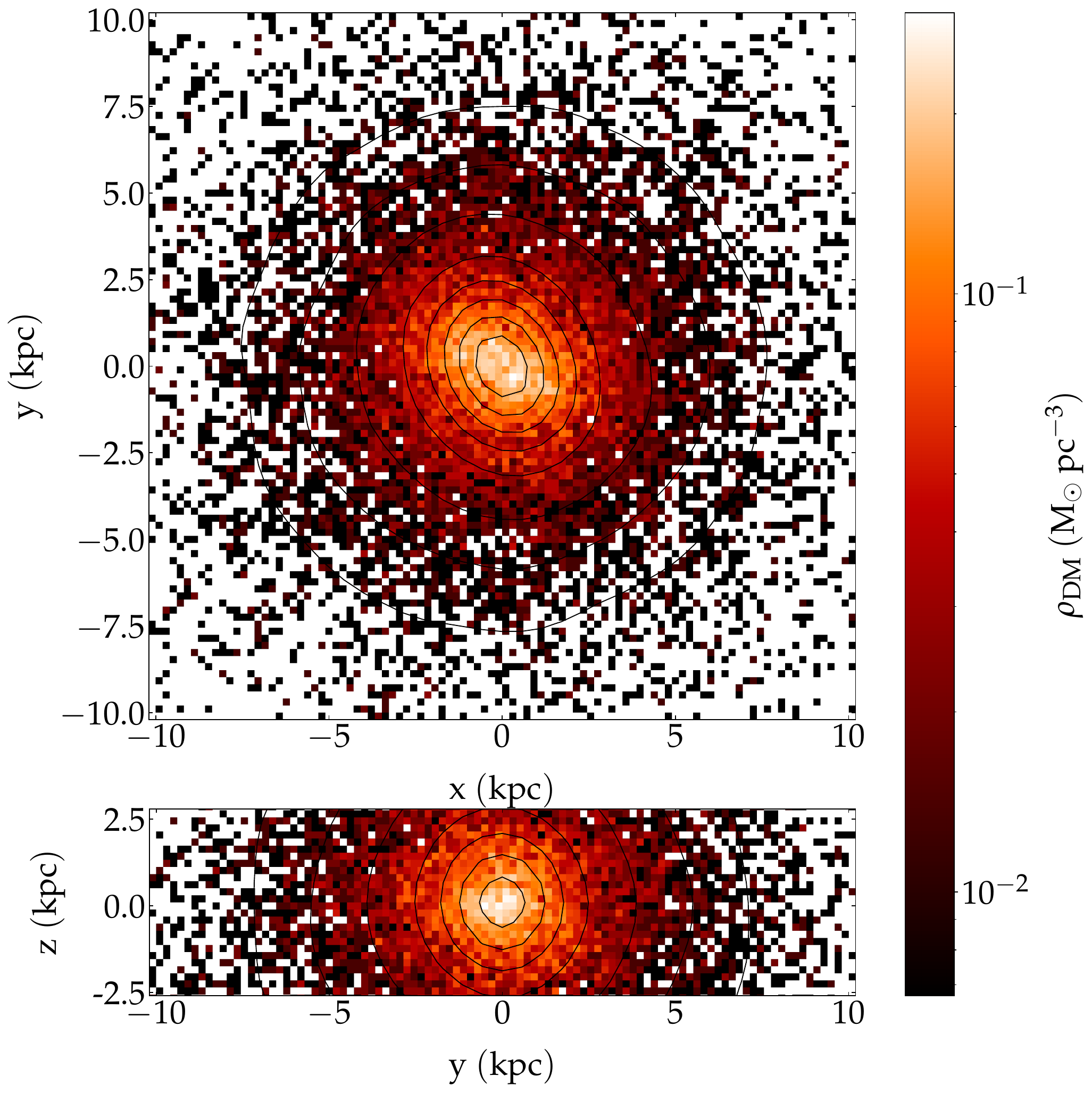}
	\caption{Comparison between the distribution of the effective dark matter, $\rho_D$ in NLG (left) and the dark matter density in the LPH model (right) at $t=590$ Myr. Top panels show the density in the $x-y$ plane and the bottom panel indicate the density in the edge-on view. The contour levels have been smoothed using the SciPy algorithm \citep{Virtanen2020}.} \label{fig:density1}
\end{figure*}

\begin{figure*}
	\centering
	\includegraphics[width=0.5\linewidth]{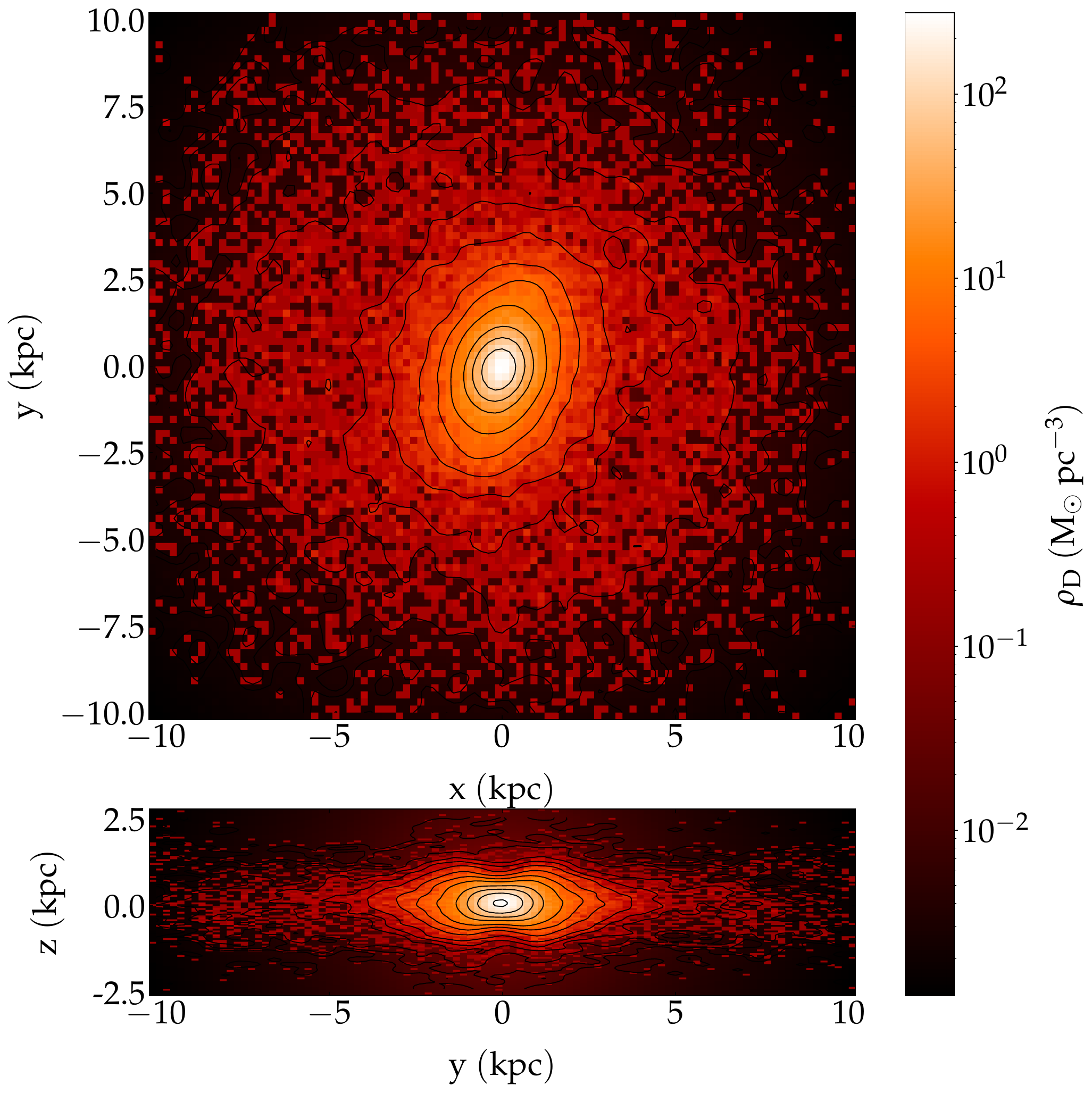}\hspace*{0.4cm}
	\includegraphics[width=0.5\linewidth]{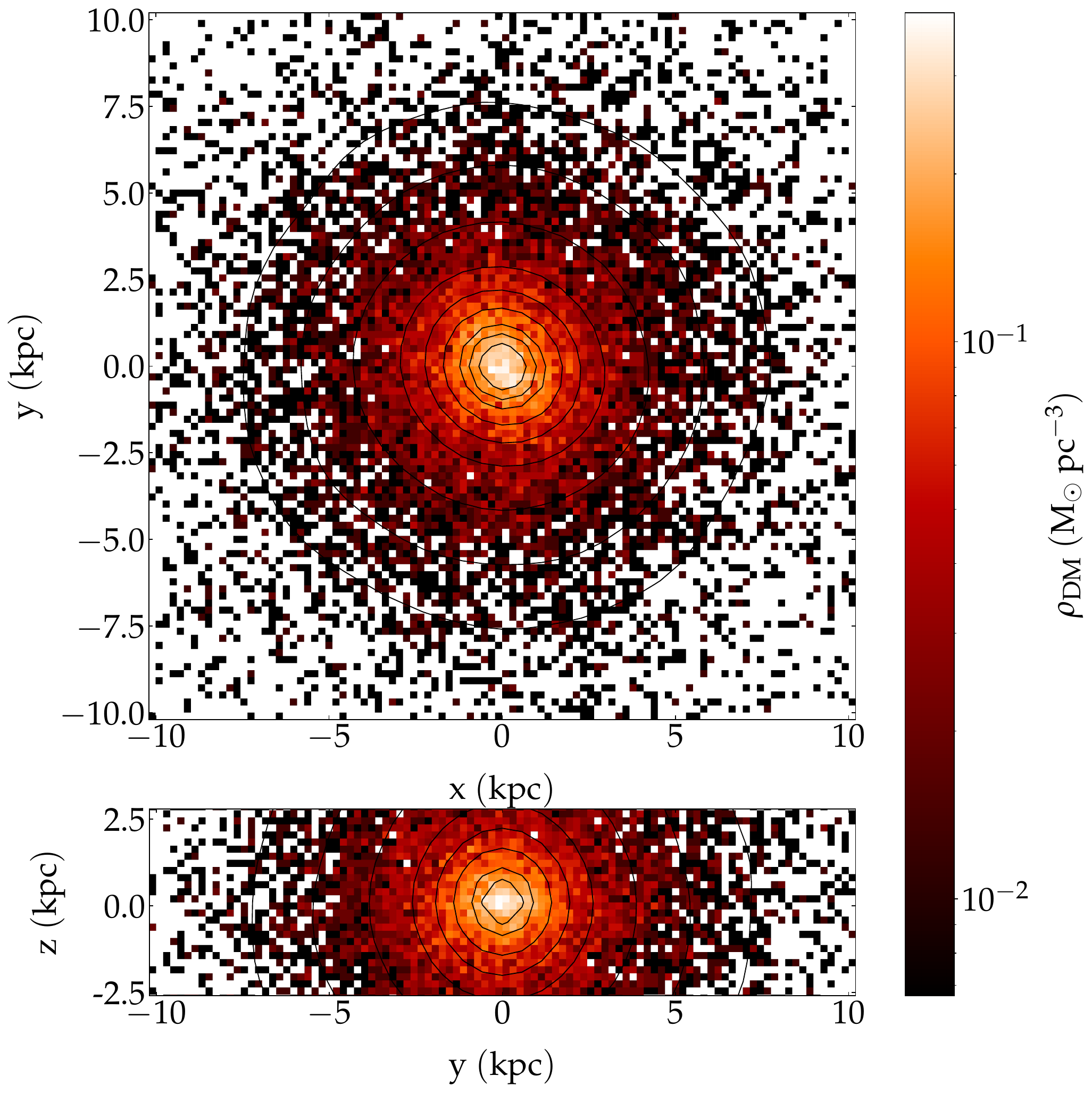}
	\caption{Similar to Figure~\ref{fig:density1} but densities are evaluated at $t=4$ Gyr.}\label{fig:density2}
\end{figure*}

By defining an arbitrary mesh grid, we estimate the effective dark matter density in the NLG model using equation \eqref{eq8} and compare it with the dark matter density in the LPH model in Figures~\ref{fig:density1} and \ref{fig:density2}.

In the NLG model, as already mentioned, the effective dark matter mimics the behaviour of the matter density. That is, if a bar or spiral structure form in the disc itself, similar structures are seen in the distribution of effective dark matter density. For instance, see the left panel in Figure~\ref{fig:density1}. This panel shows the effective dark matter density when the bar magnitude reaches its maximum value. At this time there is a spiral pattern in the baryonic disc. The same pattern appears in the effective dark matter. In contrast, in the LPH model, the dark matter halo remains almost spherical during the simulation and there is no sign of spiral structures in the dark matter density. Actually the distribution of the dark matter particles is not completely independent of the baryonic matter properties. For example, the stellar bar induce a matter wake within the dark matter halo which causes dynamical friction against the bar rotation \citep{Beane2023}. No wake can be induced in the effective dark matter distribution in NLG. This is why the dynamical friction is meaningfully smaller in NLG compared to particle dark matter models \citep{Roshan2021}.

The distribution of the effective dark matter and particle dark matter at the end of simulation are illustrated in Figure~\ref{fig:density2}. It is remarkable that there is a peanut in the effective dark matter reflecting the fact that there is a weak peanut in the baryonic matter.

The above mentioned characteristic of the effective dark matter density is the key to understanding the different behaviours of galactic discs in the NLG model.

\subsection{Unstable bending waves}
\begin{figure*}
	\centering
	\includegraphics[width=0.335\linewidth]{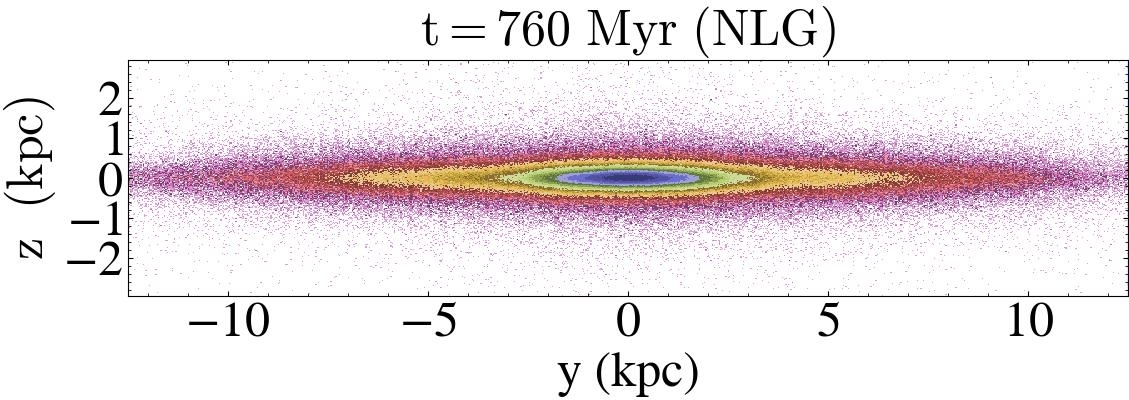}\hspace*{0.3cm}	
	\includegraphics[width=0.30\linewidth]{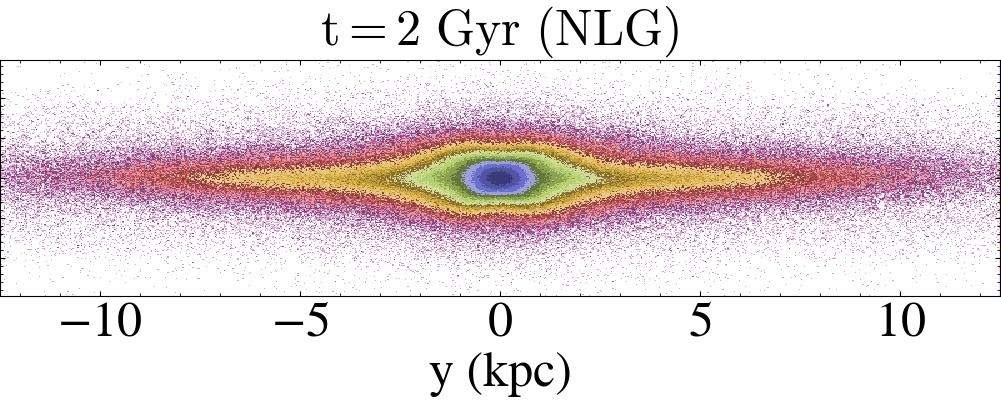}\hspace*{0.3cm}
	\includegraphics[width=0.30\linewidth]{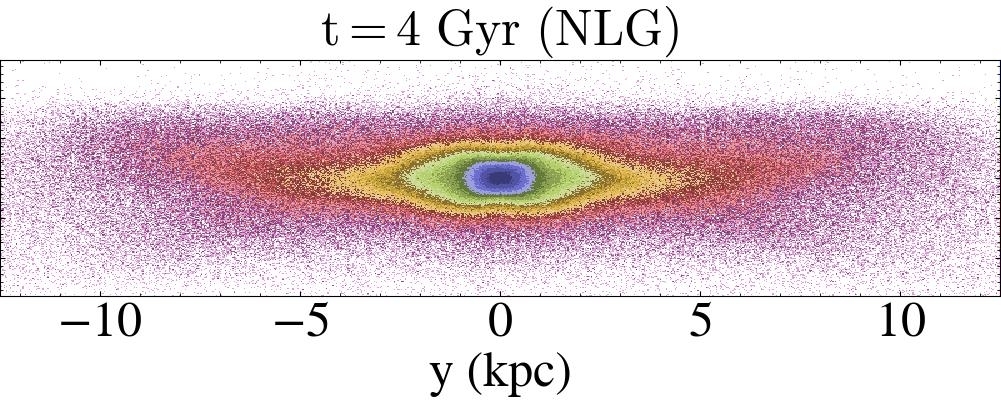}\vspace{0.5cm}
	
	\includegraphics[width=0.335\linewidth]{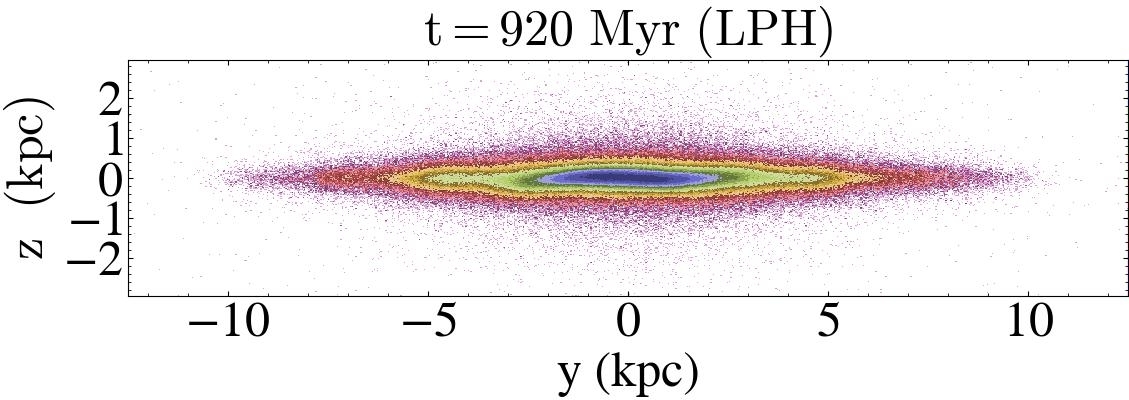}\hspace*{0.3cm}	
	\includegraphics[width=0.30\linewidth]{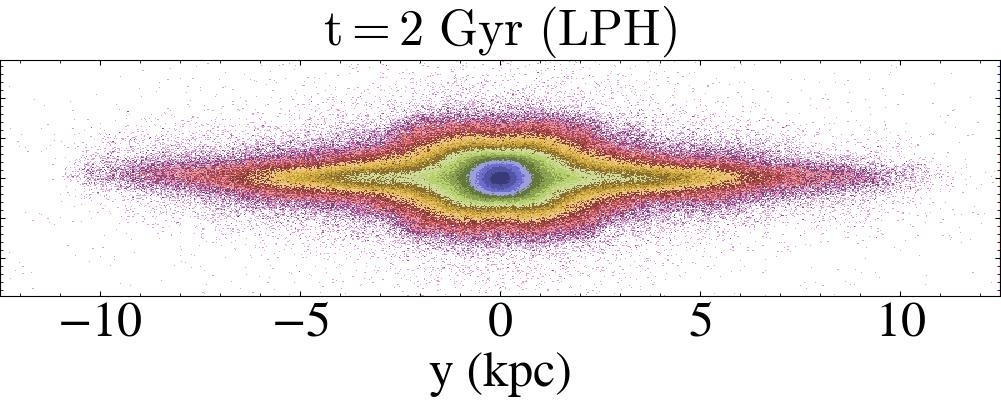}\hspace*{0.3cm}
	\includegraphics[width=0.30\linewidth]{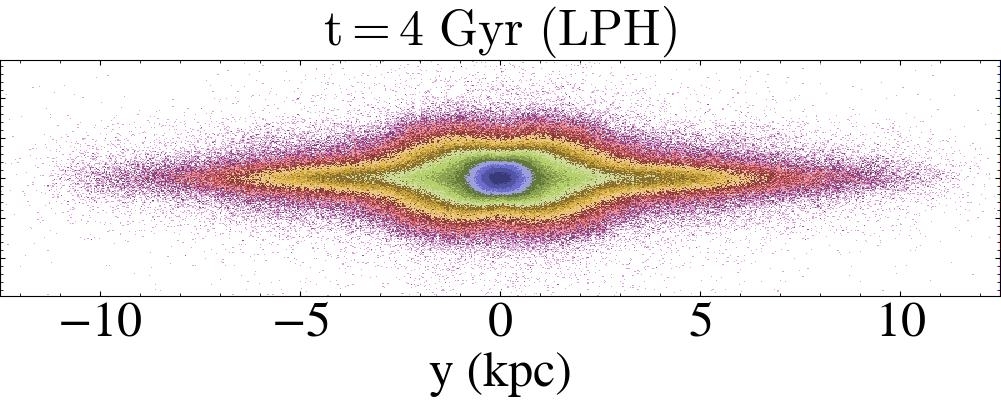}\\
	\caption{Top (bottom) panels illustrate the time evolution of the NLG (LPH) model projected in the edge-on direction.}\label{fig:bucklingLPH}
\end{figure*}

We reiterate that the vertical structure of the discs in modified gravity and dark matter models are meaningfully different. The main purpose of this paper is to shed light on the origin of this difference. Before moving on to discuss the bending instability using the relevant dispersion relations, it is useful to see the edge-on view of the discs. The vertical disc particle distributions in our models are illustrated in Figure~\ref{fig:bucklingLPH}. These edge-on views show the whole disc as it evolves from the buckling time until the end of the simulation. The top panels belong to NLG. We see that the disc gets thicker at the outer parts compared to the LPH case. On the other hand, although a strong peanut is developed in the LPH model, there is a weak peanut in the NLG model. At the end of the simulation, we see that there is a mild warp in the NLG model as well.

The manifestation of unstable bending waves in our models, is observed through the buckling of the bars and the formation of warps. In order to identify which model is more unstable against bending perturbations at a given radius, we turn our attention to all the terms in the dispersion relations \eqref{eq2} and \eqref{eq22}. In what follows in this section, we have computed all these terms from the snapshots just before the buckling time which is determined in the previous subsection.

\subsubsection{Frequency of the vertical oscillations}
As mentioned before (Section \ref{sec:analys}), $\nu_{\text{ext}}$ ($\nu_{\text{NLG}}$) is the vertical frequency due to the dark matter halo (effective dark matter in the NLG model). This quantity is determined by the second derivative of the potential of the halo with respect to $z$ or by the first derivative of the gravitational field due to all the dark matter particles. We use the latter case. So to compute the vertical frequency, we need to determine the gravitational field of the halo around the disc. To do so, we consider a Cartesian grid where the location of each mesh point gives the coordinates of the field points. Then we calculate the gravitational field caused by all the dark matter particles at each mesh point. The selected mesh size is $60\times60\times14$, and the numbers indicate the number of mesh points in the $x$, $y$, and $z$-direction, respectively. Then, we divided the disc into annuli of uniform width and computed the average gravitational field within each annulus. Finally, we take the first derivative normal to the disc, to obtain the frequency.

In order to compute the vertical frequency in the NLG model, we consider only the nonlocal correction terms in equation \eqref{eq13} for the gravitational force. Notice that this is equivalent to consider only the contribution of the effective dark matter halo. Then we follow the same procedure used in the LPH model. 

\begin{figure}
	\centering
	\includegraphics[width=1.0\linewidth]{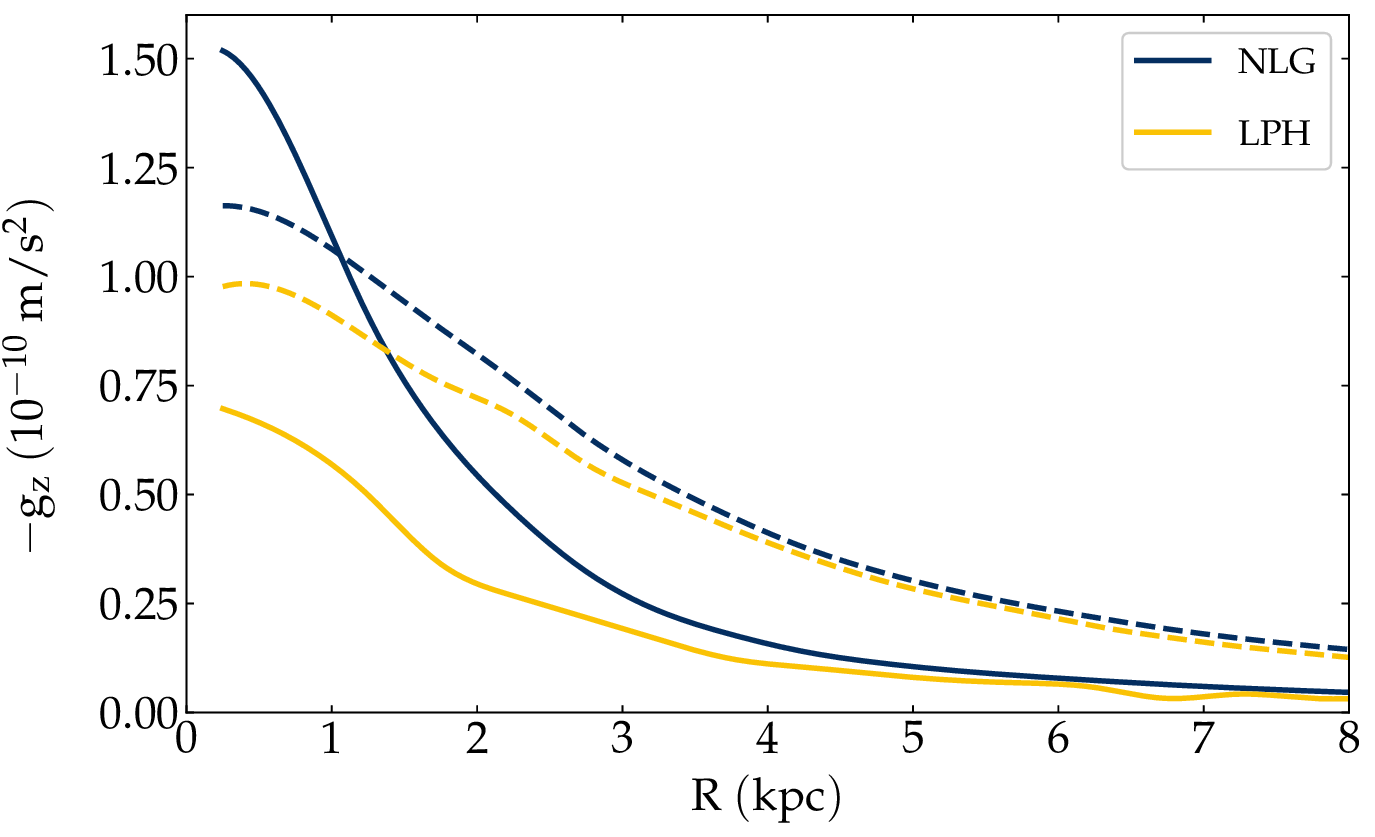}
	\caption{The $z$-component of the azimuthally averaged gravitational field in the NLG (blue) and LPH (yellow) models in terms of radius. Different planes $z=0.5$, and $z=2.0$ kpc are indicated by the solid, and dashed lines, respectively. The gravitational field is stronger at all radii in the NLG model.}\label{fig:gravity}
\end{figure}

The gravitational field in the vertical direction is illustrated in Figure~\ref{fig:gravity} for different $z$. It is interesting that the restoring force, due to the external potential, is greater in the NLG model at all radii. As mentioned before, the distribution of effective dark matter in NLG model significantly deviates from the density profile of the dark matter halo. The close relationship between the effective dark matter distribution and the density of matter which is shown by equation \eqref{eq8}, results in higher effective densities where the baryonic matter density is high. Consequently, the effective dark matter in NLG is more flattened and compressed around the disc compared to the dark matter model, see the bottom panels in Figures~\ref{fig:density1} and \ref{fig:density2}. As a result, it seems natural to expect that the vertical component of the gravitational force is greater in NLG. In other words, the gravitational force by the effective dark matter on the surface of the disc is stronger than the live particle dark matter halo in the LPH model.

 
 \begin{figure}
 	\centering
 	\includegraphics[width=1.0\linewidth]{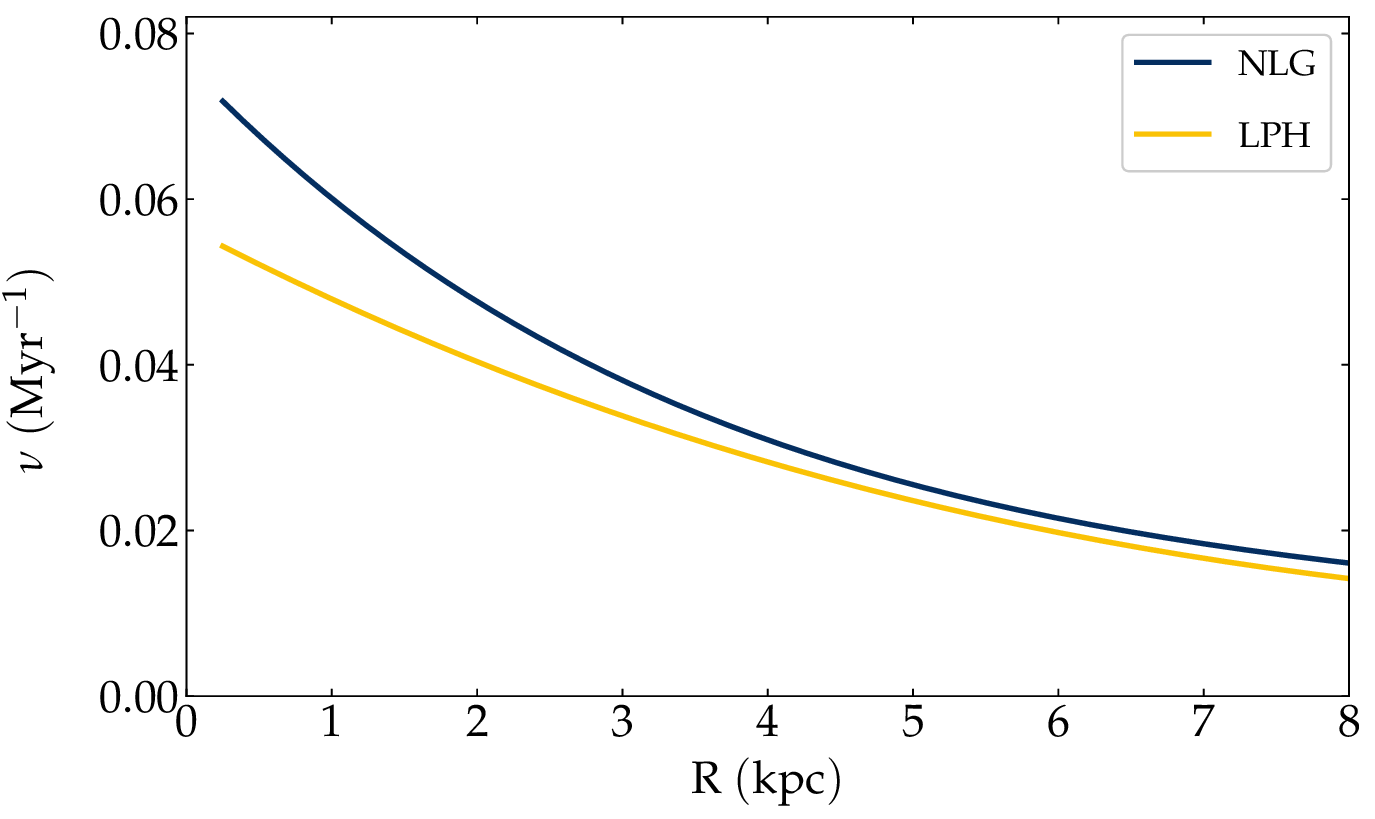}
 	\caption{The azimuthally averaged vertical frequency caused by the dark matter component (effective dark matter in the NLG model) for the NLG (blue) and LPH (yellow) models. As it is expected, the NLG model has a stronger frequency at all radii.} \label{fig:frequency}
 \end{figure}
 
Figure~\ref{fig:frequency} shows the vertical frequency in terms of radius in our models. According to the dispersion relation, the vertical frequency stabilizes the disc against bending instability. On the other hand, the vertical frequency is larger in the NLG model, specifically in the inner part of the disc. This discrepancy arises due to the distribution of effective dark matter in the NLG model, which contributes to a stronger gravitational force exerted on the surface of the disc when compared to the live particle dark matter halo in the LPH model. Therefore, we expect that the stabilizing effect of this term is more powerful in the NLG model.

\subsubsection{Self gravity of the disc}
The second terms in dispersion relations \eqref{eq2} and \eqref{eq22} are related to the self-gravity of the disc. This term has a stabilizing effect as it appears with a positive sign in the dispersion relation and strengthens the restoring force.

The self-gravity of the disc avoids vertical distortions in much the same way as elasticity in conventional membranes. In a disc with small thickness, this term can provide vertical cohesion and guarantees the adiabatic invariance of the vertical oscillations of stars as they follow the bend \citep{Merritt1994, Sellwood1998, Debattista1999}.

\begin{figure}
	\centering
	\includegraphics[width=1\linewidth]{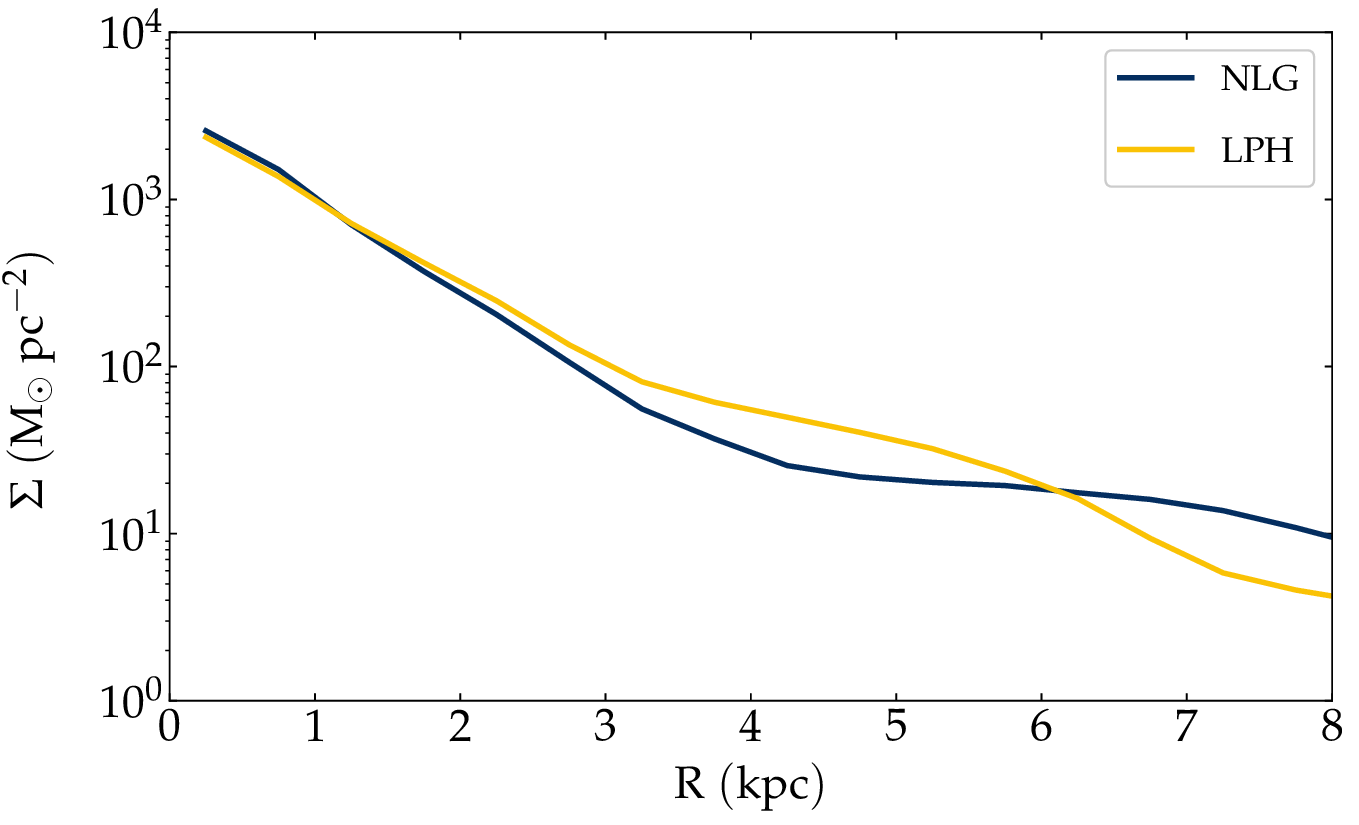}
	\caption{The azimuthally averaged surface density in terms of radius for the NLG (blue) and the LPH (yellow) models at the buckling time. Densities are shown on a logarithmic scale. \label{fig:sigma}}
\end{figure}

Obviously, for a given wave number $k$, the self-gravity is provided by the surface density of the disc. To compute this quantity, we adopted a method where the disc is divided into annuli of equal width, characterized by mean cylindrical radius $R$. We then sum up the number of particles within each annulus to calculate the surface density as a function of radius. This approach allowed us to obtain the azimuthally averaged surface density in practice. Figure~\ref{fig:sigma} shows the radial evolution of the surface density for the NLG and LPH models at the buckling time.

We see that in the innermost regions the stellar density is the same in both models. On the other hand, in the middle radii, the stellar density is higher in the LPH model. It should be noted that the radius at which $M(R)=0.95 M_{\text{disc}}$ is $R=6$ kpc in the LPH model, while for NLG, this radius is almost equal to $R=9$ kpc. This means that, the radial extension  of the NLG disc to the larger radii results in higher stellar density in the outer parts compared to the LPH model. .

\subsubsection{The velocity dispersion} 
Since the only destabilizing term in the dispersion relation includes the horizontal velocity dispersion, this quantity plays a key role in the bending instability in the disc. In each snapshot, the dispersion velocities are computed by dividing the disc into annuli of equal width and evaluating velocity dispersion components at each annulus. Specifically, the vertical velocity dispersion $\sigma_z(R)$ can be obtained using the relation $\sigma_z=\sqrt{\langle v_z^2\rangle-\langle v_z\rangle^2}$, which $\langle\rangle$ denotes the averaging over all of the particles within each annulus. It should be noted that, all the dispersion velocities are calculated at the mid-plane of the disc. As mentioned before, this means that we restricted our analysis to particles confined within $|z|\leq 1$ kpc, and projected all these particles onto the $z=0$ plane.

\begin{figure}
	\centering
	\includegraphics[width=1.0\linewidth]{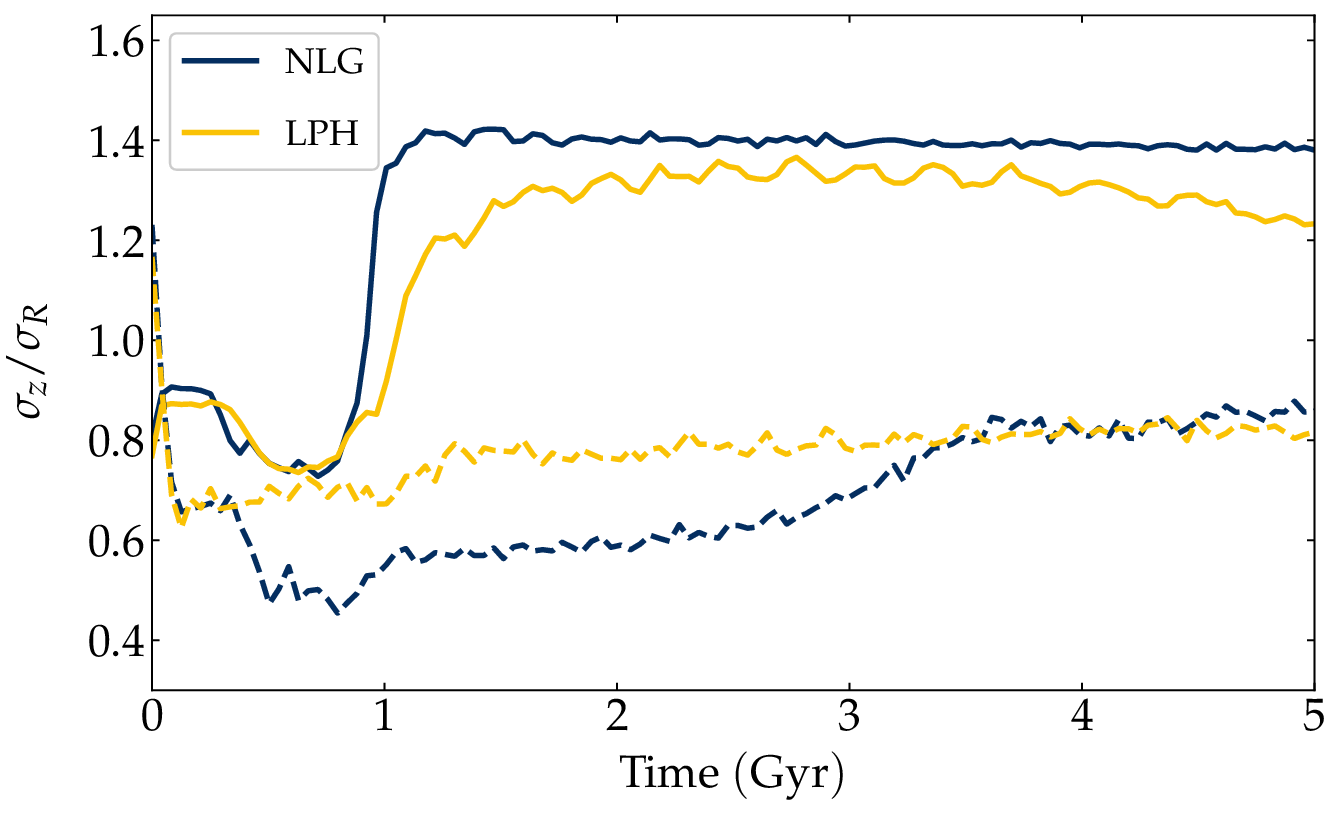}
	\caption{Time evolution of the ratio $\sigma_z/\sigma_R$ at $R=1$ kpc for the NLG (blue) and LPH (yellow) models. The dashed lines indicate this ratio at the outer regions $R=5$ kpc.\label{fig:sigma2_t}}
\end{figure}

During the bar formation in the disc, the radial dispersion velocity of the stars grows more quickly than the vertical one. So one can expect a drop in the ratio $\sigma_z/\sigma_R$ right after the bar formation. The temporal variation of the ratio $\sigma_z/\sigma_R$ is illustrated in Figure~\ref{fig:sigma2_t}. The solid lines indicate the time evolution of $\sigma_z/\sigma_R$ at $R=1$ kpc. As expected, we see a dip in this ratio right after the bar formation in each model. After the buckling instability, the vertical dispersion velocity increases more rapidly in the NLG model, which is a signature of buckling strength in this model. As illustrated by dashed lines, the ratio $\sigma_z/\sigma_R$ does not change effectively at the outer radius of the LPH model ($R=5$ kpc) during the buckling instability. So one can conclude that the NLG model is more unstable in the outer regions.  

\begin{figure}
	\centering
	\includegraphics[width=1.0\linewidth]{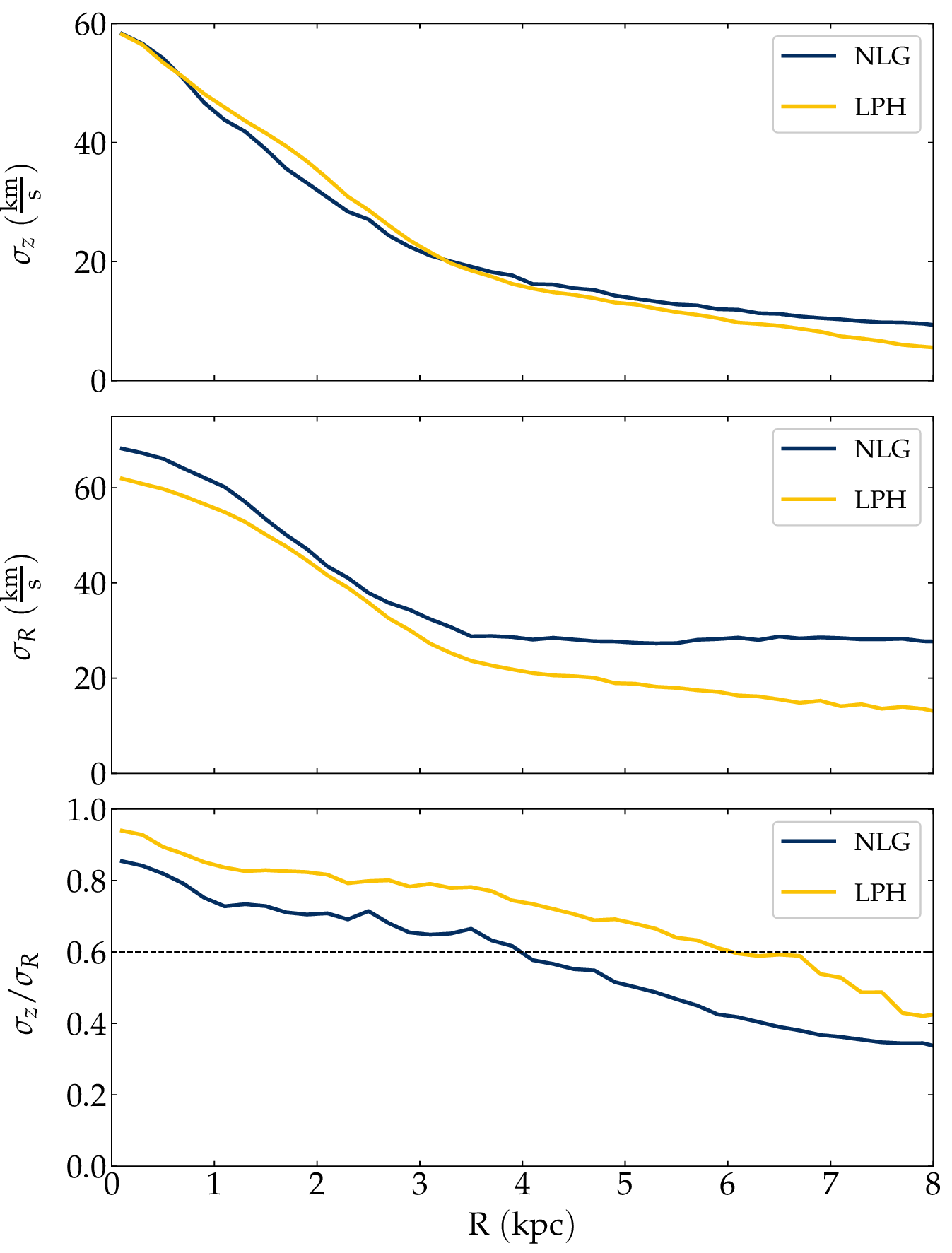}
	\caption{\textit{Top panel}: evolution of the vertical velocity dispersion in terms of radius for the NLG (blue) and LPH (yellow) models. \textit{Middle panel}:  the radial dispersion velocity as a function of radius for our models. \textit{Bottom panel}: the ratio of the velocity dispersion in the vertical direction $\sigma_z$ to the radial velocity dispersion $\sigma_R$. The dashed line indicates the criterion suggested by \citep{Merritt1994}. The measurements are made at buckling time.\label{fig:sigma2}}
\end{figure}

Figure~\ref{fig:sigma2} shows the evolution of velocity dispersions $\sigma_z$ and $\sigma_R$ in terms of radius for the NLG and LPH models at the buckling time. The middle panel of this figure clearly shows that the radial dispersion velocity $\sigma_R$ is higher in the NLG model, especially in the outer regions of the disc where the relative difference exceeds 100\%. This difference in dispersion velocity makes the outer parts of the disc more unstable in the NLG model, as the other quantities in equations \eqref{eq2} and \eqref{eq22} do not exhibit significant differences in the outer radii, as illustrated by Figures~\ref{fig:frequency} and \ref{fig:sigma}.
So the question is: why is the radial dispersion higher in NLG? One of the primary factors contributing to the increase in radial dispersion velocity is the presence of spiral patterns. The heating process resulting from these spirals induces a secular rise in the velocity dispersion of the stars. \citep{Sellwood1984, Carlberg1985}. As already mentioned, the increase in radial dispersion velocity renders the disc prone to bending instability.

\begin{figure*}
	\centering
	\includegraphics[width=0.269\linewidth]{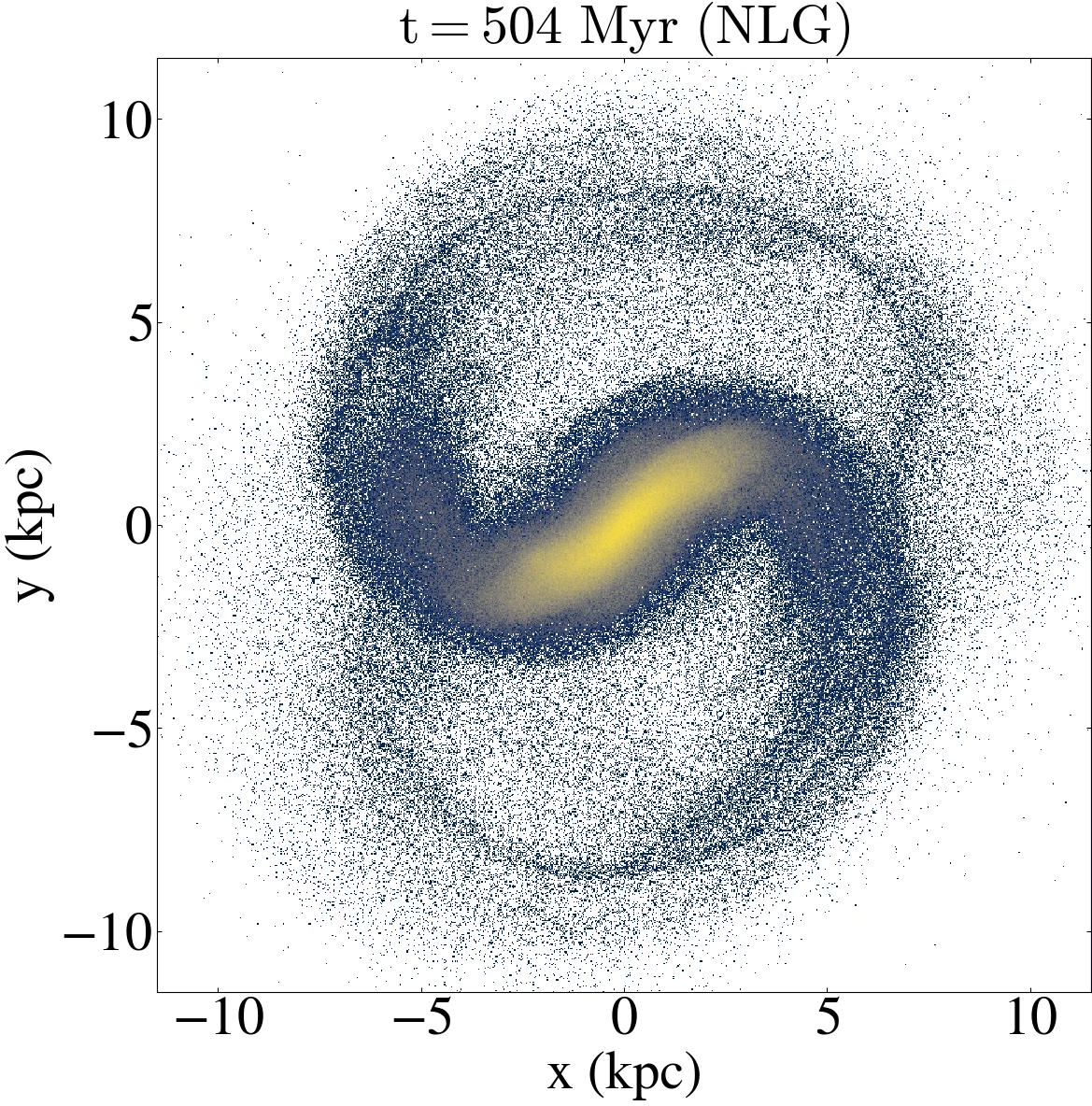}	
	\includegraphics[width=0.23\linewidth]{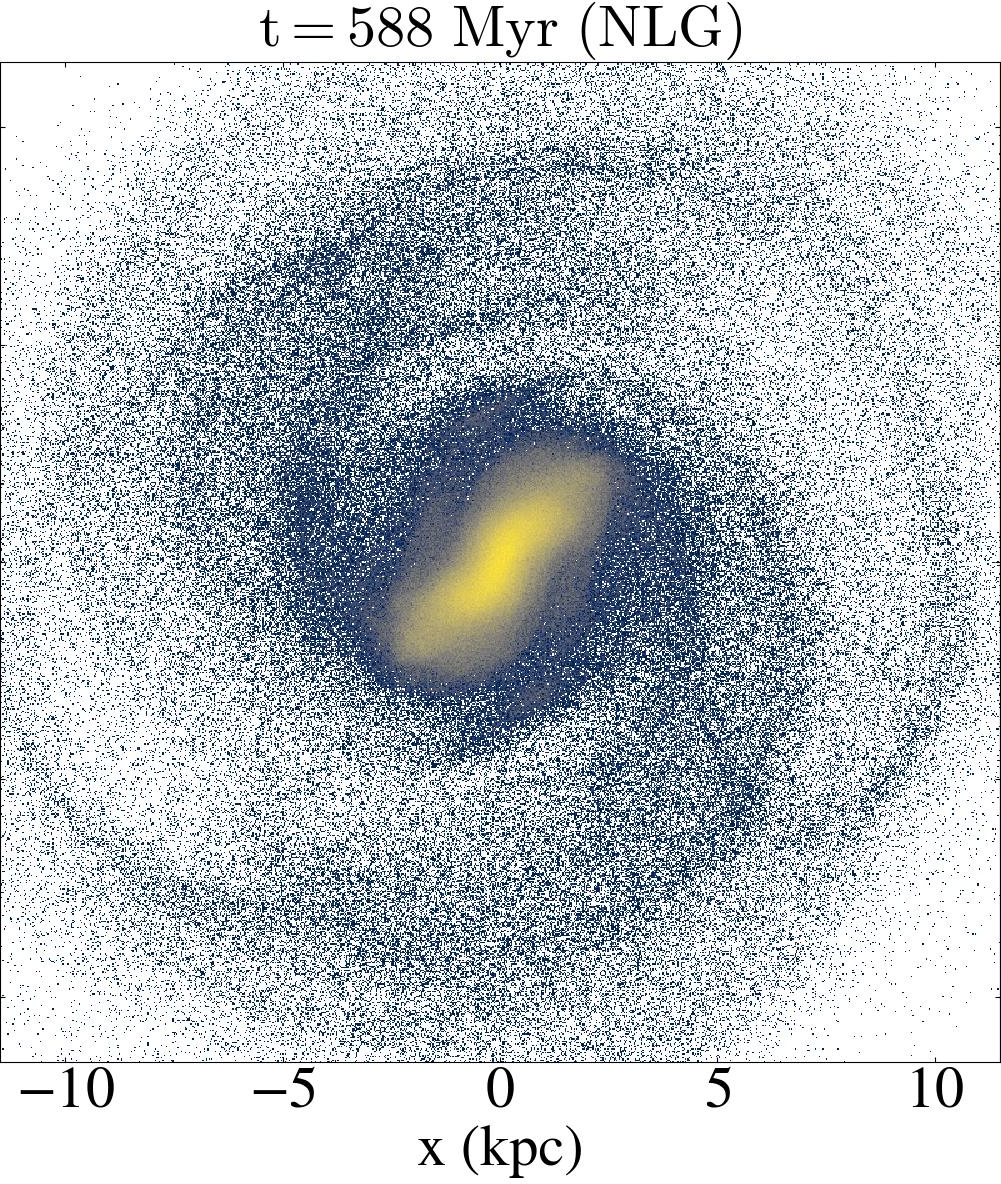}
	\includegraphics[width=0.23\linewidth]{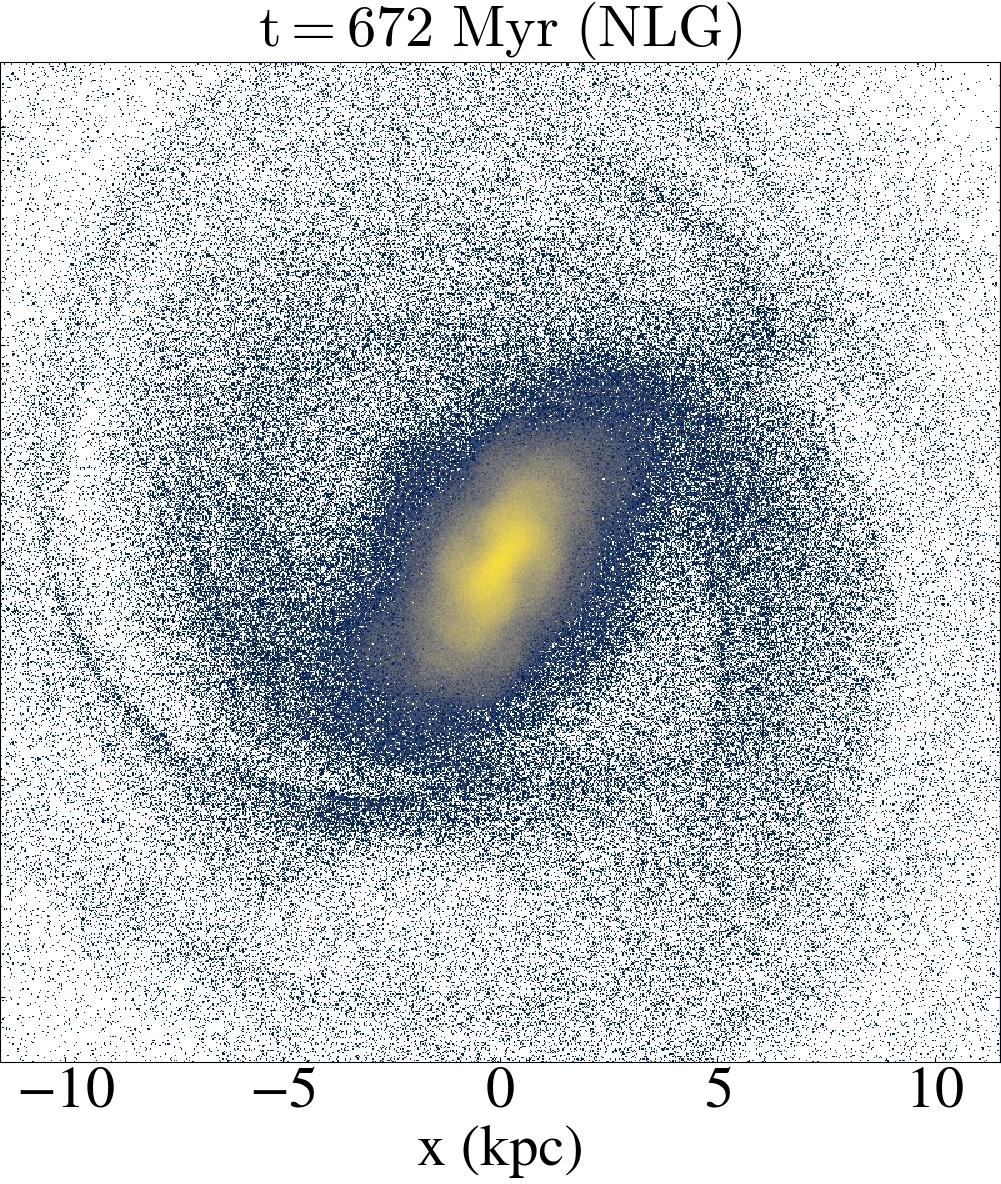}
	\includegraphics[width=0.23\linewidth]{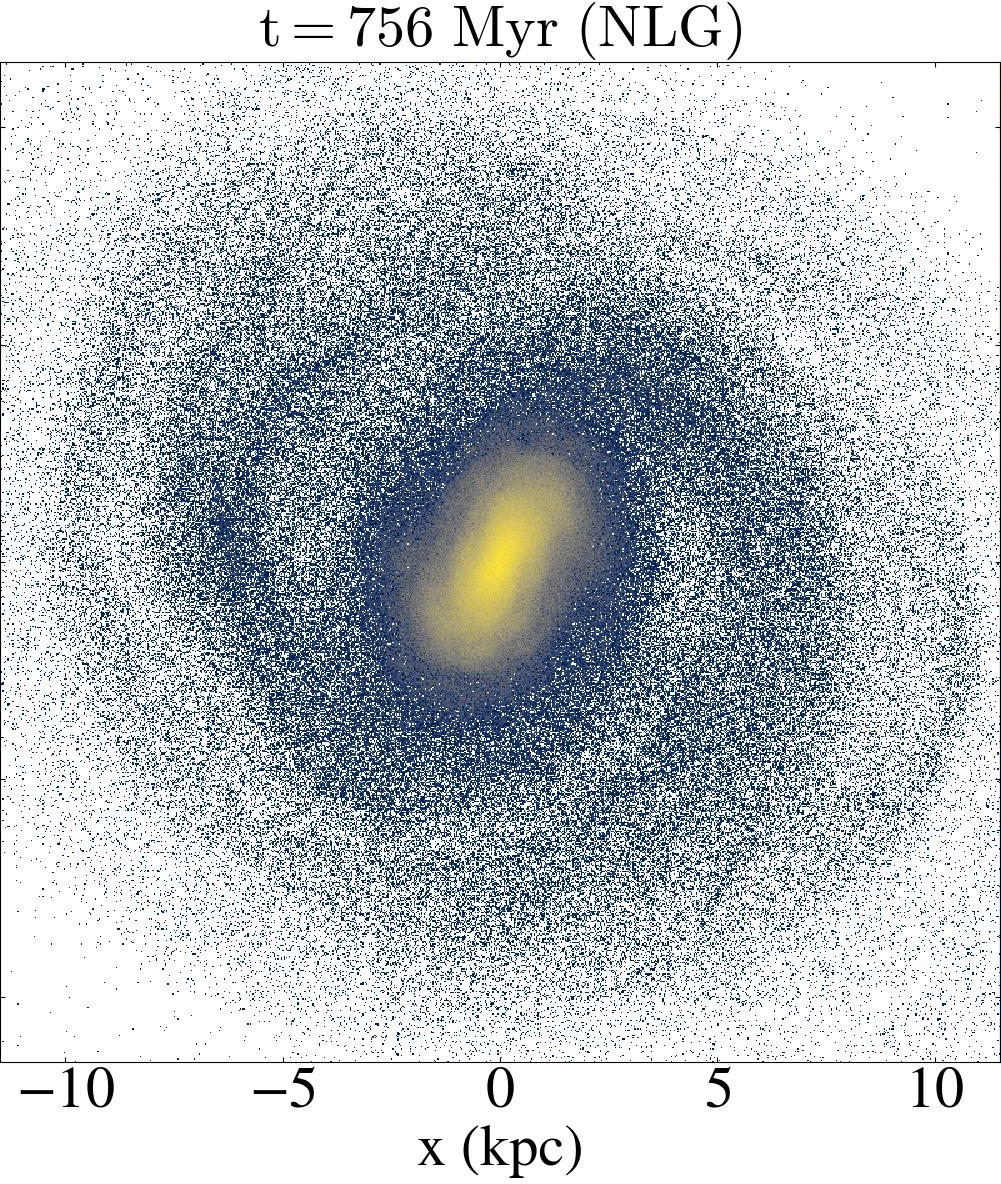}\vspace*{0.5cm}
	\centering
	\includegraphics[width=0.2685\linewidth]{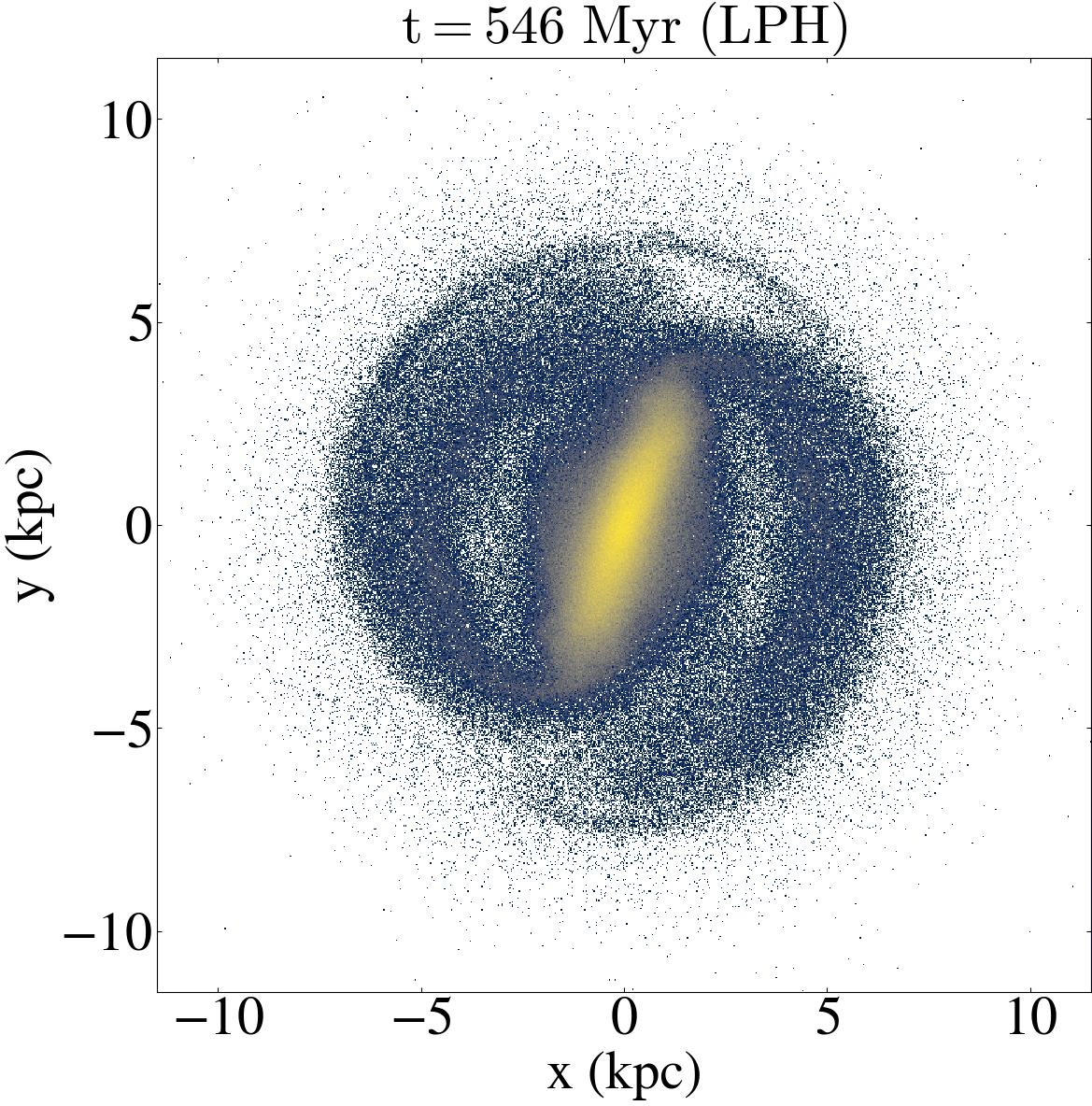}	
	\includegraphics[width=0.23\linewidth]{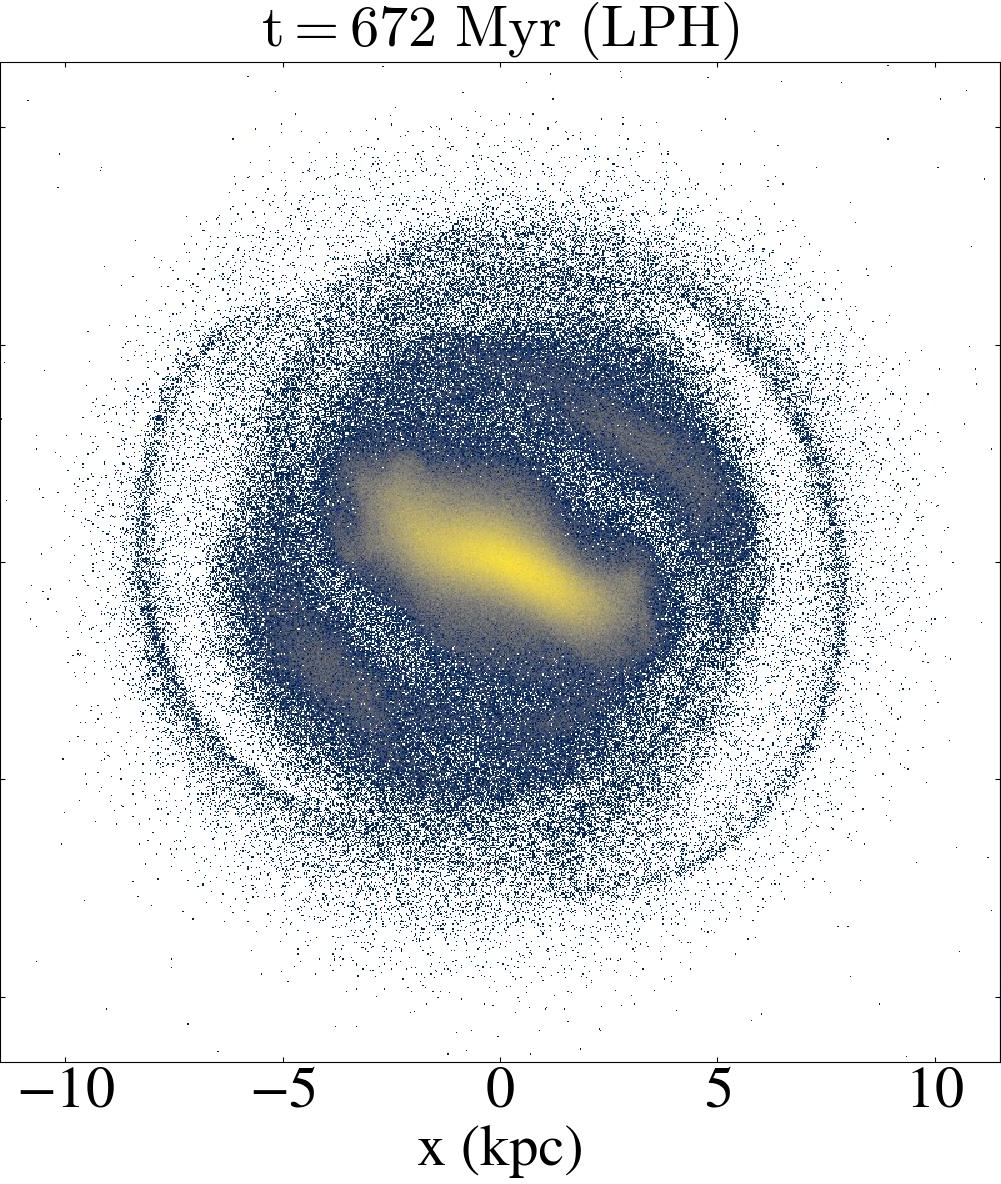}
	\includegraphics[width=0.23\linewidth]{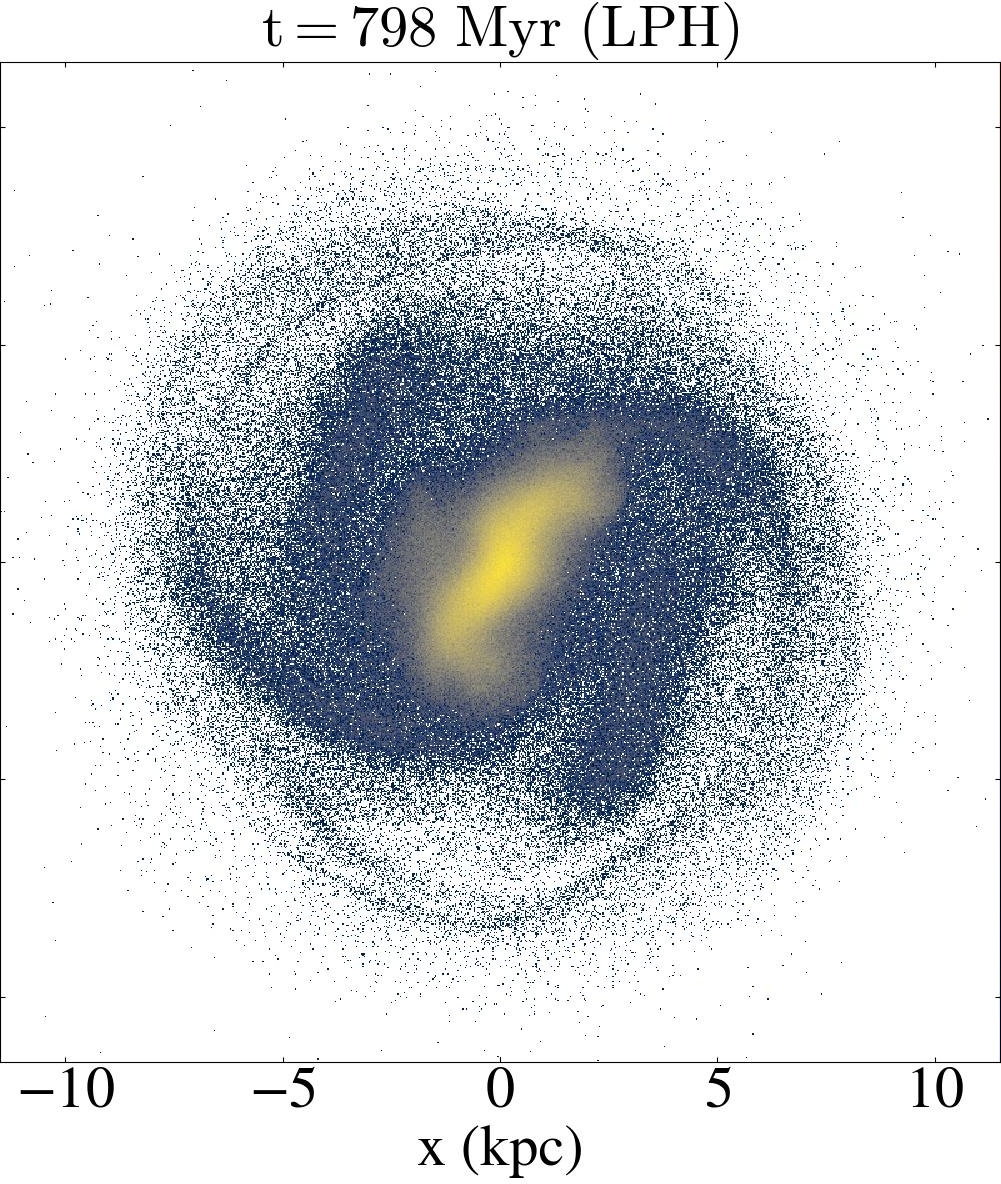}
	\includegraphics[width=0.23\linewidth]{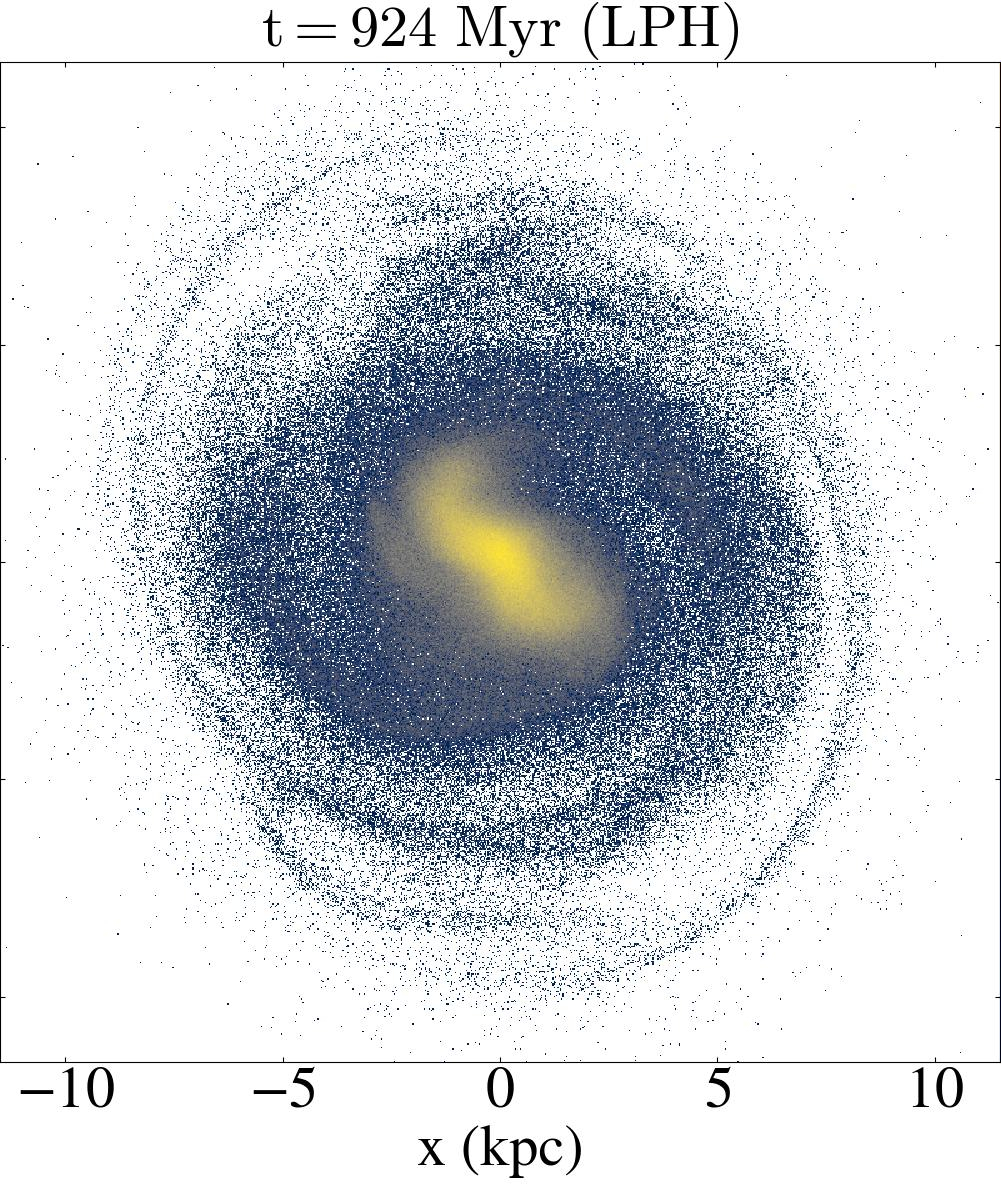}
	\caption{Top (bottom) panels illustrate the time evolution of the NLG (LPH) model projected in the $x-y$ plane.}\label{fig:face_on}
\end{figure*}

As depicted in Figure~\ref{fig:face_on}, the NLG model demonstrates more pronounced and extended spiral patterns, reaching larger radii compared to the LPH model. The figure showcases the face-on views of the disc in both the NLG (top panels) and LPH models (bottom panels) from the onset of bar formation until just prior to the onset of the buckling instability. This relatively short time interval is sufficient for the bar/spiral activity to redistribute angular momentum throughout the disc. By computing the pattern speeds in our models, it is evident that the bar/spiral in the NLG model rotates approximately 8 times within this interval, whereas the LPH model completes nearly 12 rotations during this period. Spirals in NLG propagate to larger radii resulting in higher radial velocity dispersion in these regions as indicated in Figure \ref{fig:sigma2}.  

The ratio $\sigma_z/\sigma_R$ as a function of $R$ is also illustrated in the bottom panel of Figure~\ref{fig:sigma2}. For the two models, dispersion in the vertical direction drops more effectively than the radial velocity dispersion. So one expects smaller values for the local ratio $\sigma_z/\sigma_R$ at the outer regions.

It has been shown that the disc is locally unstable against bending instability if $\sigma_z/\sigma_R \lesssim 0.29-0.37$ \citep{Toomre1966, Kulsrud1971, Araki1985}. Although, \citet{Merritt1994} suggested that $\sigma_z/\sigma_R \lesssim 0.6$ is a more accurate criterion for a realistic axisymmetric disc. It is interesting that both of our models satisfy this criterion in the outer regions. It should be mentioned that the smaller ratio of $\sigma_z/\sigma_R$ at the buckling time provided the stronger pressure of the buckling force \citep{Collier2020}. So one may expect that the outer regions of the disc would be more prone to instability in the NLG model.

It is worth mentioning that \citet{Sotnikova2003} suggested that the bending instability in the central regions of a stellar disc is completely suppressed at $\sigma_z/\sigma_R \approx 0.75-0.8$.

\subsubsection{Unstable wavelengths}
As we mentioned before, dispersion relations \eqref{eq2} and \eqref{eq22} are derived for a razor-thin disc. It should be noted that when we study the bending instability in a disc with finite thickness, these relations are applicable only for perturbations with wavelengths longer than the thickness of the disc \citep{Toomre1966, Sotnikova2005, Binney2008}. 

In our simulations, the disc has a vertical density profile of the form $\text{sech}^2(z/2z_d)$ where $z_d$ is the vertical scale height of the disc. By considering $z_d$ as a measure of the thickness, we analyzed the bending instability in our models for the perturbations with $1\leqslant\lambda\leqslant25$, where $\lambda = 2\pi/k$ is the perturbation wavelength in terms of kpc.

\begin{figure*}
	\centering
	\includegraphics[width=0.995\linewidth]{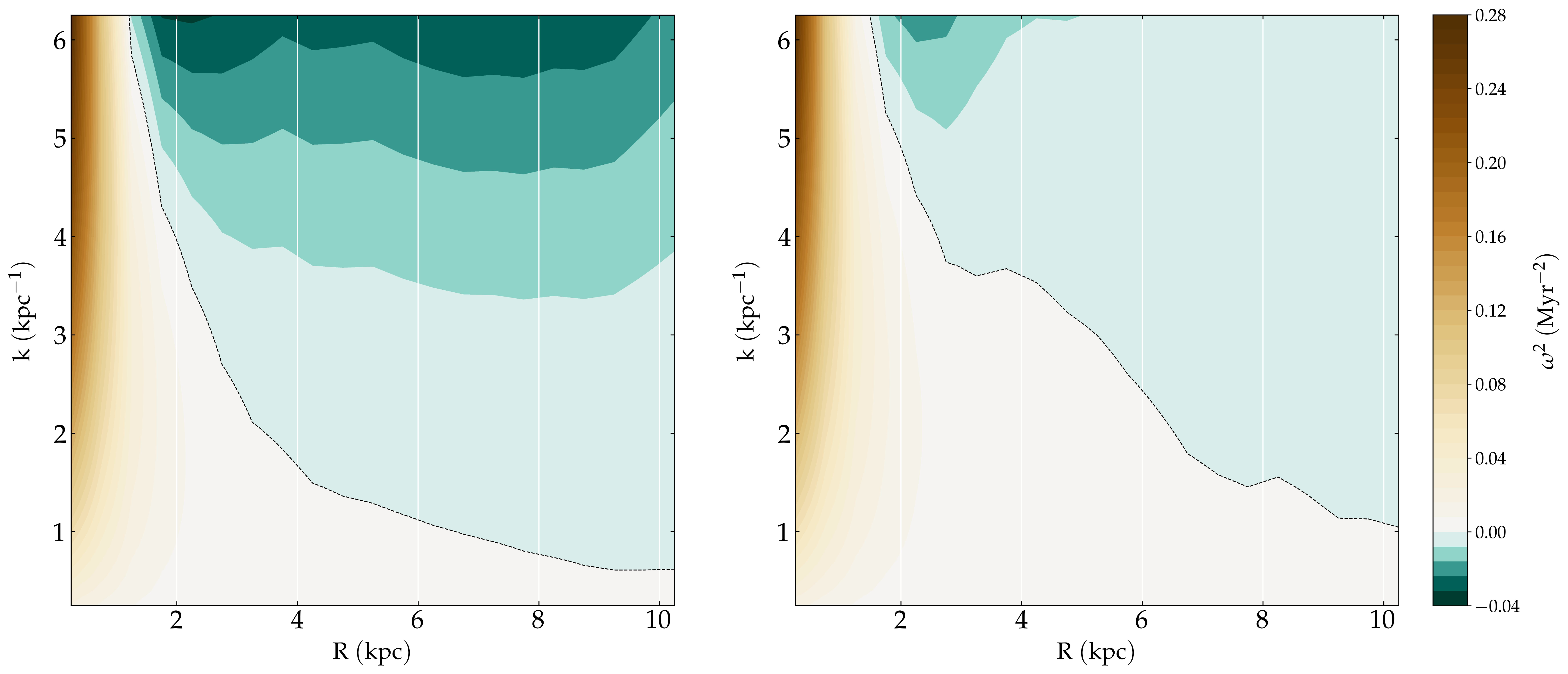}\\
	\includegraphics[width=0.976\linewidth]{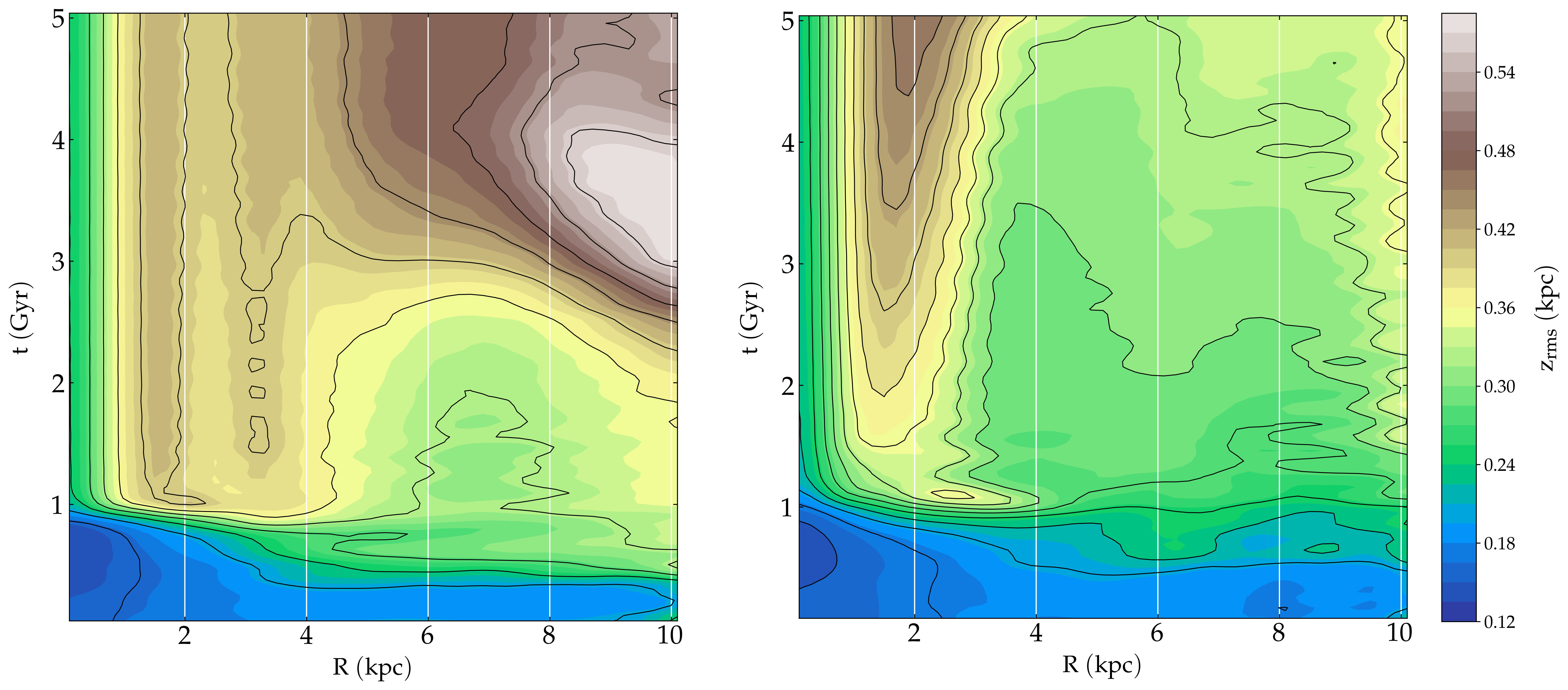}
	\caption{\textit{Top panels}: contour plots of $\omega^2$ in the $R-k$ plane for the NLG (left) and LPH (right) models. The contour levels are  indicated by colours. \textit{Bottom panels}: the root-mean-square thickness, $z_{\text{rms}}$, at different radii in terms of time for the NLG (left) and LPH (right) models. The contours of $z_{\text{rms}}$ have been smoothed using the SciPy algorithms \citep{Virtanen2020}.\label{fig:omega}}
\end{figure*}

The results of our analysis are illustrated in Figure~\ref{fig:omega}. In the top panels, we show contour plots of $\omega^2(R,k)$ for the NLG (left panel) and LPH (right panel) models in the $R-k$ plane. The contour levels are indicated by colours. 

Both of our models are stable against long-wavelength perturbations, namely $k<0.5\,\text{kpc}^{-1}$. However, at a given radius the disc becomes unstable by decreasing the wavelength. As illustrated in Figure~\ref{fig:omega}, bending instability is more violent in the NLG model at almost all the regions of the disc. We expect that the disc becomes thicker at the large radii in the NLG model as the instability rate is higher compared to the LPH case. On the other hand, the LPH model is more unstable in the inner regions with $1\leqslant R\leqslant3$, for the short wavelengths, which indicates that the peanut shapes are more expected in these regions. We can see that for both models the region $R\gtrsim 1$ kpc is unstable to short-wavelength perturbations according to our dispersion relations. However the growth rate of the perturbations are meaningfully higher in the NLG model specially at outer regions. To be specific, the growth rate $|\omega|$ at $R=6.25$ kpc is $|\omega|= 0.17\ 1/\text{Myr}$ and $|\omega|= 0.09\ 1/\text{Myr}$ in the NLG and LPH models respectively. So, in comparison with the dynamical time scale in our models, that is $\tau_0=4.2\ \text{Myr}$, the instability time scale $\tau\approx 1/|\omega|$ in the NLG model is $\tau \simeq 1.4\tau_0$ whereas in the LPH model $\tau \simeq 2.7\tau_0$. It means that the instability occurs more slowly in the LPH model. It should be noted that the growth rate in the LPH model is at most $|\omega|= 0.14\ 1/\text{Myr}$ at $R=2.25$ kpc so it seems that in overall the disc is more unstable in the NLG model. It is also necessary to mention that the instability region in the top panels of Figure~\ref{fig:omega} is more extended in the NLG model and covers a wider range of unstable wavelengths. Notice that the dashed curve shows the border of stability and instability zones. Therefore, not only the growth rates are higher in NLG, but the instability region is wider in NLG. We recall that all these interpretations are extracted from a single snapshot right before the onset of instability.

Although it is not possible to infer from the top panels of the Figure~\ref{fig:omega} and equivalently from our dispersion relations, that the final peanut is stronger in LPH, it is interesting that only a small region in the $\omega^2(R,k)$ panel has the highest growth rate in the LPH model. More interestingly, this region projected on the $R$ axis is close to the position of the peanut lobes. 
\subsection{The impacts of the bending instability}

A rapid increase in the scale height of the disc and the formation of a boxy/peanut-shaped structure are the expected consequences of bending instability in the inner part of the disc. However, based on the unstable bending waves, the effect of this instability could be different in the context of modified gravity and dark matter models. Several numerical studies have shown that in the inner parts of the disc, the modified gravity models predict a smaller thickness, while in the dark matter model, discs seem to be thicker and show stronger peanut-shaped structures. On the other hand, galactic discs in modified gravity models show higher thickness at the large radii \citep{Ghafourian2020, Roshan2021}. As we discussed, from the top panels of Figure~\ref{fig:omega}, we also confirm that the NLG model is more prone to bending instability at the outer parts of the disc. However, our analysis was based only on a single snapshot. For having a more complete picture, we need to consider the time evolution of the bending instability. Therefore, in order to investigate how the bending instability affects the different parts of the discs in our models, we trace the evolution of thickness using the root-mean-square vertical height, $z_{\text{rms}}$. The bottom panels in Figure~\ref{fig:omega} display $z_{\text{rms}}$ for the particles with $|z|\leq 10 z_d$ at several radii as a function of time. 

Following the occurrence of the bending instability ($t\geqslant1$ Gyr), the thickness of the disc increases in both models; however, the structure of the disc is different. In the NLG model, there is no sign of strong peanut-shaped structures in the inner parts, and the disc is thicker at the outer radii. While in the LPH model, the disc becomes thicker at a limited region in the inner parts ($1\leqslant R\leqslant3$), and the thickness of the outer regions does not significantly change over time. It should be noted that the brown region in the lower right panel of Figure~\ref{fig:omega} indicates that the peanut-shaped structures are developed in the inner parts of the disc in the LPH model. These results are consistent with our expectations from the analysis of the propagation of the bending waves encoded in the top panels of Figure~\ref{fig:omega}. 

The long-term evolution of our models reveals distinct differences in the final structure of the galactic disc. In the NLG model, the strength of the bar remains relatively constant throughout the evolution, showing minimal variation over time. On the other hand, the LPH model exhibits a continuous increase in bar strength. Notably, the NLG model produces a final disc configuration that is more radially extended compared to the LPH model. Furthermore, as the long-term evolution progresses, the NLG model sustains a thicker disc structure compared to the LPH model. The disc in the NLG model maintains a greater vertical extent, while the characteristic peanut-shaped structure, often associated with the presence of a strong bar, tends to dissipate and become less prominent in the NLG model as the evolution continues.

\section{Conclusions}\label{sec:con}
In this paper, we investigated the propagation of bending waves in the vertical direction of galactic discs in NLG and compared it with the standard dark matter model. Following an approximated model, we derive a dispersion relation for the bending waves in the vertical direction of discs in NLG. Then we explored the predictions made by this analytical approach through the high-resolution N-body simulation. 

In the presence of a bending perturbation, the stars in the disc are forced to vertically oscillate when they encounter the bending wave. The bending instability grows when the restoring force of the disc is not enough to provide the required vertical acceleration. According to the dispersion relation, the restoring force depends on the surface density of the disc and the gravitational force due to the external potential. In a dark matter model, this external potential is caused by a dark matter halo, while in the NLG model, the nonlocal terms of gravity provide this potential. 

At a given wave number, the stability of the disc against the bending perturbation is provided by the vertical frequency due to the external potential and the surface density of the disc. However, there is a destabilizing term in the dispersion relation, which includes the horizontal dispersion velocity of the stars. In order to investigate which model is more unstable against this instability, we constructed two models in NLG and the standard dark matter model with a live Plummer halo (LPH) using the N-body simulations and measured the above-mentioned quantities at the onset of the bending instability, where it manifests as the buckling instability. We showed that, the vertical frequency $\nu_{\text{NLG}}$, caused by the effective dark matter halo in the NLG model, is larger than the vertical frequency $\nu$ in the LPH model at all radii. The observed discrepancy can be attributed to the rather flattened distribution of effective dark matter in the NLG model, which results in a more pronounced gravitational force exerted on the surface  of the disc compared to the spherical dark matter halo in the LPH model. On the other hand, the NLG model exhibits a higher radial dispersion velocity, particularly at larger radii. This can be attributed to the presence of stronger and more extended spiral arms in the NLG model. The heating process induced by these spirals leads to a secular increase in the velocity dispersion of the stars. It should be noted the effect of the surface density is not so determinant because it depends only on the disc component, and surface density is not so different in the two models. 

One can easily compute the frequency $\omega$ of a propagating bending wave for several wave numbers at different radii. As illustrated in Figure~\ref{fig:omega}, we expect the disc in the NLG model to be more unstable, especially in the outer regions. It is worth mentioning that we consider only the perturbations with wavelengths larger than the scale height $z_d$ of the disc because our discs have a finite thickness, whereas the dispersion relations are derived for a razor-thin disc. 

We trace the evolution of the vertical structure of the disc using the root-mean-square $z_{\text{rms}}$. Although the disc in both of our models becomes thicker after the occurrence of the instability, the overall structure of the disc is different in our models. In the LPH model, the inner parts are thicker, while the NLG model exhibits a disc which is thickened at the outer parts. We conclude that a bending wave with a given wave number propagates in different ways in a disc embedded in a dark matter halo and under the effect of the nonlocal gravity model. This phenomenon can explain the different behaviour of the galactic discs in the vertical direction in these two paradigms. The origin for this difference can be simply attributed to the different distribution of the particle dark matter and the effective dark matter. This fact directly affects the vertical structure of galactic discs.

To summarise our findings, the main results are highlighted below.
\begin{enumerate}
	\item The growth rate of the bending perturbations is significantly higher in the NLG model, particularly in the outer regions. This indicates that the instability occurs at a slower pace in the LPH model.
	\item The instability region is more extended in the NLG model, encompassing a broader range of unstable wavelengths compared to the LPH model.
	\item In the NLG model, the thickness of the outer parts undergoes a significant increase over time. In contrast, the LPH model shows an increased thickness within a limited region in the inner parts.
	\item In the NLG model, there is no sign of strong peanut-shaped structures; however, the presence of warped discs is more anticipated. Conversely, the LPH model exhibits a distinct and prominent peanut-shaped structure.
\end{enumerate}

Given the significant disparities between the structure of galactic discs in the NLG and LPH models, an important question arises as to whether these differences can be discerned through observational investigations. In the following we discuss why answering this question at the moment is quite challenging.

The instability time scale is considered as one potential probe; however, it is important to note the limitations in accessing the time evolution of real galaxies. Moreover, previous studies have shown that the number of observed currently buckling barred galaxies is quite limited \citep{Erwin2016, Li2017, Xiang2021}. Such observational limitations poses a significant challenge in studying the time scale of the instability in real galactic discs.

While examining the consistency of each theory through simulation for a specific galaxy is a viable approach, it is crucial to accept the limitations of isolated N-body simulations in directly comparing them with observational data. These limitations stem from the significant impact of various factors, including the presence of a gas component, on the vertical structure of galactic discs. Notably, the inclusion of gas is known to play a pivotal role in the formation of warps within dark matter simulations \citep{Revaz2004}. Consequently, to effectively distinguish between these two theories, a case study of a specific galaxy utilizing hydrodynamical simulations becomes essential, as it allows for a more comprehensive examination of the interplay between gas dynamics and the vertical structure of galactic discs.

Indeed, statistical studies conducted on observational samples provide another avenue to distinguish between the theories. For instance, examining the percentage of peanut structures or warps in a sample of edge-on galaxies can be highly informative. However, it is important to emphasize that achieving a quantitative comparison with such observational samples requires the utilization of a large sample of simulated galaxies obtained from cosmological simulations within both paradigms. Currently, there are no cosmological simulations within the framework of modified gravity theories that reject the existence of dark matter particles, whereas there are many cosmological simulations within the standard cold dark matter model.	

Our study does not incorporate a gas component and also cannot yield to a statistical analysis. Consequently, it lacks a firm basis for comparison with real data. However, it can still serve as an initial step and highlight potential differences between nonlocal gravity and the cold dark matter model. The subsequent step would involve performing hydrodynamical isolated simulations in NLG (or in modified gravity in general).

It should be noted that, although our paper focuses on the unstable bending waves in the vertical direction, it is necessary to consider the role of the vertical resonance in the buckling instability, which is responsible for the formation of boxy/peanut shapes in galactic discs. Previous studies have highlighted the importance of the vertical resonance in this process, particularly in the context of dark matter model~\citep{Combes1990, Quillen2002, Quillen2014, Saha2018}. However, further studies are necessary to fully understand the role of the vertical resonance in the vertical structure of galactic discs in modified gravity models. In particular, it would be interesting to determine whether the vertical resonance can account for the differences between the vertical structure of galactic discs in dark matter and modified gravity paradigms.

\section*{Acknowledgements}
This research made use of the Sci-HPC centre of Ferdowsi University of Mashhad. We would like to appreciate the referee, Aneesh P. Naik, for his instructive comments that helped us a lot to improve this work. TK is grateful to Benoit Famaey for useful
discussions.  This work is supported by Ferdowsi University of Mashhad under Grant No. 56145 (13/9/1400).
\section*{Data Availability}
The data presented in this work can be obtained upon reasonable request.

\bibliographystyle{mnras}
\bibliography{vertical_structure} 

\appendix

%

\bsp	
\label{lastpage}
\end{document}